\documentclass[twocolumn,preprintnumbers, amssymb,amsmath,aps,floatfix,prd,nofootinbib,superscriptaddress,showpacs]{revtex4-1}

\usepackage{epsfig}
\usepackage{bm}
\usepackage{amssymb}
\usepackage{amsmath}
\usepackage{color}
\usepackage{subfigure}
\usepackage[colorlinks,
            linkcolor=blue,
            anchorcolor=black,
            citecolor=blue
            ]{hyperref}

\begin{document}
\title{Lepton Pair Production Through Two Photon Process in Heavy Ion Collisions}

\author{Spencer Klein}
\affiliation{Nuclear Science Division, Lawrence Berkeley National
Laboratory, Berkeley, CA 94720, USA}

\author{A. H. Mueller}
\affiliation{Department of Physics, Columbia University, New York, NY 10027, USA}

\author{Bo-Wen Xiao}
\affiliation{Key Laboratory of Quark and Lepton Physics (MOE) and Institute
of Particle Physics, Central China Normal University, Wuhan 430079, China}

\author{Feng Yuan}
\affiliation{Nuclear Science Division, Lawrence Berkeley National
Laboratory, Berkeley, CA 94720, USA}

\date{\today}

\begin{abstract}
This paper investigates the electromagnetic production of lepton pairs with low transverse momentum in relativistic heavy ion collisions.
We estimate the initial photons' transverse momentum contributions by employing models where the average transverse momentum squared of the incoming photon can be calculated in the equivalent photon approximation. We further derive an all order QED resummation for the soft photon radiation, which gives an excellent description of the ATLAS data in ultra-peripheral collisions at the LHC. For peripheral and central collisions, additional $p_T$-broadening effects from multiple interaction with the medium and the magnetic field contributions from the quark-gluon plasma are also discussed.
\end{abstract}
\maketitle

\section{Introduction}

There have been strong interests on the electromagnetic process of dilepton production in two photon scattering in heavy ion collisions in the last few years~\cite{Aaboud:2018eph,Adam:2018tdm,Lehner:2019amb,Adam:2019mby,ATLAS:2019vxg}. These experiments have made great efforts to select the dilepton events through the kinematic constraints where they are produced by the purely electromagnetic two-photon reaction: $\gamma\gamma\to\ell^+\ell^-$. These experimental results have attracted quite a lot of attentions in the community because they may provide a potential probe for the electromagnetic property of the quark-gluon plasma created in heavy ion collisions. The effects being considered include the medium induced transverse momentum ($p_T$) broadening from multiple interaction effects and/or the magnetic field effects from the medium. The confirmation of either effect would be an important pathway in future explorations of the electromagnetic properties of the quark-gluon plasma. 

Furthermore, the physics related to the acoplanarity of the dilepton pair is very similar to that in dijet productions described in QCD. One popular description of the jet medium interactions in QCD is the Baier-Dokshitzer-Mueller-Peigne-Schiff (BDMPS) approach~\cite{Baier:1996kr, Baier:1996sk, Baier:1998kq}. In addition to jet energy loss, the BDMPS formalism also predicts the $p_T$ broadening effects, since these two effects are physically related. In the BDMPS formalism, $\hat{q} L$ characterizes the typical transverse momentum squared that a parton acquires in the medium of length $L$. The experimental study of the medium related $p_T$-broadening effects of jet is an important step forward to clarify and understand the underlying mechanism for the jet energy loss in heavy ion collisions. In the last few years, there have been a lot of progresses in the understanding of the $p_T$-broadening effects in dijet, photon-jet, and hadron-jet productions in heavy ion collisions~\cite{Mueller:2016gko,{Mueller:2016xoc},{Chen:2016vem},{Chen:2016jfu},{Chen:2018fqu},Tannenbaum:2017afg}. It was found the vacuum Sudakov effects dominates~\cite{Mueller:2016gko} the transverse momentum broadening of dijets  for the typical dijet kinematics measured at the LHC~\cite{Aad:2010bu,Chatrchyan:2011sx}. On the other hand, in the RHIC energy regime, the quark-gluon-plasma medium effect is comparable to the Sudakov effects, and the medium $p_T$-broadening has been observed in the measurements of hadron-jet correlation~\cite{Adamczyk:2017yhe,Chen:2016vem} by the STAR collaboration. We expect future measurements at both the LHC and RHIC will allow us to gain further important and quantitative information on the medium transverse momentum broadening effect. 

To be able to probe in-medium or magnetic effects in peripheral collisions, it is important to have a baseline measurement. Ultra-peripheral collisions (UPCs), where the nuclei do not interact hadronically (roughly, with impact parameter $b>2R_A$ where $R_A$ is the nuclear radius), can provide this baseline data. UPCs encompass both photo-production and two-photon interactions \cite{Bertulani:1987tz,Baur:2007fv,Bertulani:2005ru,Baltz:2007kq, Harland-Lang:2018iur} and have a long history. Using the so-called equivalent photon approximation\cite{Fermi:1924tc}, which is also known as the Weizs\"acker Williams method \cite{vonWeizsacker:1934nji, Williams:1934ad}, the photon distribution from a relativistically moving charge particle can be computed from its boosted electromagnetic fields. Photons can strike the oppositely moving nucleus, or they can interact with the electromagnetic field of the other nucleus, creating a pair of leptons in the two-photon process $\gamma\gamma\to\mu^+\mu^-$. In two-photon fusion processes, the final state has a small total $p_T$, so the leptons are nearly back-to-back.

The first dedicated study of $\gamma\gamma\rightarrow e^+e^-$ was by the STAR Collaboration in 2004 \cite{Adams:2004rz}. STAR studied final states consisting of the lepton pair, accompanied by neutrons in each zero degree calorimeter. The neutrons are assumed to come from the mutual Coulomb excitations of the two nuclei by two additional photons, not directly associated with the pair production \cite{Baltz:2009jk}. The additional photons biased the production towards smaller impact parameters (but still with $b>2R_A$)\cite{Baur:2003ar}.   

Two-photon fusion production of lepton pairs has been calculated with both the equivalent-photon approximation (EPA), where the photons are treated as massless \cite{Baltz:2009jk}, and via full lowest-order QED calculations, which included non-zero virtual photon masses \cite{Hencken:2004td}.  The two calculations agreed well, except that the EPA calculation predicted more pairs with pair $p_T < 20$ MeV/c. The data was in good agreement with the calculations, favoring the QED calculation at small pair $p_T$. The STAR paper attributed this to a breakdown of the EPA approach, but the issue may have been that the EPA calculation did not account for the change in single photon transverse momentum ($k_T$) distribution due to the biasing toward small impact parameters \cite{Zha:2018tlq}. This can be seen as broadening the pair $p_T$ spectrum.  

Newer UPC studies of dileptons have mostly focused on $J/\psi$ or heavier vector mesons, with the two-photon production treated as a background.  However, the ATLAS collaboration has recently made a high-statistics study of two-photon production of muon pairs in UPCs, finding mostly good agreement with an EPA calculation, but accompanied by a tail containing about 1\% of the events, of pairs with relatively high acoplanarity \cite{ATLAS:2016vdy}. Acoplanarity is closely related to the pair $p_T$.  ATLAS focused on acoplanarity to reduce their sensitivity to resolution effects.

At the same time, several experiments have reported on the two-photon production of lepton pairs in peripheral collisions, with $b<2R_A$. The ATLAS collaboration studied pair production in lead-lead collisions and observed a significant increase in acoplanarity with decreasing centrality, consistent with an increase in pair $p_T$~\cite{Aaboud:2018eph}.  The STAR Collaboration studied pair production at low $p_T,$ $p_T<$ 200 MeV/c in gold-gold and uranium-uranium collisions \cite{Adam:2018tdm}, also found evidence for momentum broadening, with $p_T$ spectra different from calculations without in-medium effects \cite{Klein:2016yzr, Klein:2018cjh,Zha:2018ywo, Klein:1999qj}. This broadening could be attributed to in-medium scattering of the produced leptons, but first it is necessary to account for possible changes to the $p_T$ spectrum due to changes in the impact-parameter distribution, due to the requirements of either additional photon exchange or hadronic interactions. 

To better understand these developments from STAR and ATLAS, we study the acoplanarity of lepton pairs in heavy ion collisions. We extend our previous studies of dijet azimuthal correlations in heavy ion collisions to di-lepton angular correlations. Because the experimentally measured lepton pairs~\cite{{Aaboud:2018eph}} have very small pair $p_T$, the associated physics is a bit different than for dijet correlation. We focus on three important aspects here. The first one is the transverse momentum distribution of the incoming photons in the two-photon processes. In particular, these distributions should depend on the impact parameter of the collisions. However, as far as we know, there is no first-principle calculation on the joint transverse momentum and impact parameter dependent photon distribution within the original EPA framework. On one hand, we first make some approximation and estimate the average transverse momentum for the photons, which turns out a mild dependence on the impact parameter. On the other hand, we try to generalize the EPA framework by introducing the photon Wigner distribution which contains both the transverse momentum and impact parameter information of the incoming photon, and find that this generalization indicates the impact parameter dependence of the transverse momentum broadening could be stronger in the central collisions. We would like to emphasize that further theoretical developments are needed to address this issue. 

The second one is the QED Sudakov effects in the lepton pair production. Much of this study will be similar to the previous studies for dijet azimuthal correlations. However,
even beyond the much smaller QED coupling constant, the QED Sudakov has its own unique features. The formalism is much simpler, and more importantly, the Sudakov contribution has distinguishable behavior compared to the primordial $k_T$ distribution from the two incoming photon fluxes. The theory predictions for the UPC events agree very well with recent measurement from ATLAS~\cite{ATLAS:2016vdy}, which, for the first time, clearly demonstrates the importance of the Sudakov effects in the moderately larger acoplanarity region. Third, we discuss the QED medium effects on the pair $p_T$-broadening due to the leptons. This part is similar to the BDMPS formalism. With this physics included, we compare our calculation to the experimental data and comment on the implications of the ATLAS measurements.

It is quite interesting to compare the $p_T$-broadening effects in QED and in QCD, which helps to provide a new perspective of studying the property of quark gluon plasma created in heavy ion collisions. The medium $p_T$-broadening effect of lepton pairs is the probe to the electromagnetic constituents of the quark-gluon plasma, whereas the QCD jet $p_T$-broadening effect measures the strong interaction property. The experimental and theoretical investigations of both phenomena will deepen our understanding of the hot medium created in these collisions. More importantly, the lepton $p_T$-broadening effects can be clearly seen in the measurements of ATLAS and STAR~\cite{Aaboud:2018eph,{Adam:2018tdm}}. This is in contrast to the jet $p_T$-broadening effects in the measurement of dijet azimuthal angle correlations, due to strong QCD parton showers~\cite{Mueller:2016gko,{Mueller:2016xoc},{Chen:2016vem},{Chen:2016jfu},{Chen:2018fqu}}.  

A brief summary of our results has been published earlier in Ref.~\cite{Klein:2018fmp}. The rest of the paper is organized as follows. In Sec.~II, we discuss the $\gamma\gamma\rightarrow \ell^+\ell^-$ process in heavy ion collisions. In Sec. III,  we derive the Sudakov resummation and also compare to recent ATLAS measurement on the azimuthal correlation of the lepton pair in the UPC events and demonstrate the importance of the Sudakov contribution to the lepton pair production in these QED processes. In Sec.~IV, we derive the medium effects on the $p_T$-broadening of leptons. In particular, we compare to the QCD processes in the BDMPS formalism, and argue that the effects observed by the ATLAS collaboration are consistent with the parametric estimate of the $p_T$-broadening effects for the QED and QCD processes. Finally, after discussing possible magnetic effects in Sec. ~V, we summarize our paper in Sec. VI. 

\section{Leading Order Picture}

\begin{figure}
\includegraphics[width=8cm]{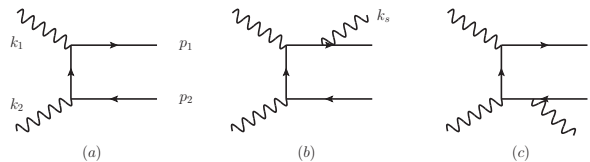}
\caption{The leading order and next-to-leading order QED Feynman graphs for the lepton pair production in two photon fusion processes.}
\label{fig1}
\end{figure}

As shown in Fig.~\ref{fig1}(a), the leading order production of lepton pairs comes from the photon-photon fusion scattering, which is the main ingredient in the STARLIGHT simulation~\cite{{Klein:2016yzr},Klein:2018cjh}. The total cross section for $\gamma+\gamma\to\mu^++\mu^-$ has a so-called $t$-channel singularity, and one has to include the lepton mass to regulate the divergence. However, we are interested in high mass di-muon production with large transverse momentum for each lepton~\cite{ATLAS:2016vdy}, where the $t$-channel singularity is absent.

Let us specify the kinematics by setting the momenta of outgoing leptons to be $p_1$ and $p_2$ with individual transverse momenta $p_{1T}$ and $p_{2T}$, and rapidities to $y_1$ and $y_2$, respectively. The leptons are produced dominantly back-to-back in the transverse plane, with $|\vec{p}_T|=|\vec{p}_{1T}+\vec{p}_{2T}|\ll |p_{1T}|\sim  |p_{2T}|$. Then, neglecting the lepton masses, we find the following longitudinal momenta for the incoming photon
\begin{eqnarray}
p_1&=&\frac{P_{T}}{\sqrt{s}}\left(e^{y_1}+e^{y_2}\right)P_A\ ,\\
p_2&=&\frac{P_{T}}{\sqrt{s}}\left(e^{-y_1}+e^{-y_2}\right)P_B\ ,
\end{eqnarray}
where $P_T =\frac{1}{2}|\vec{p}_{1T}-\vec{p}_{2T}|\sim |p_{1T}|\sim |p_{2T}|$ is much greater than 
the total momentum $p_T$, $P_A$ and $P_B$ are the incident nucleus momenta (per nucleon), respectively. The leading order differential cross section can be written as
\begin{eqnarray}
\frac{d\sigma(AB[\gamma\gamma]\to\mu^+\mu^-)}{dy_1dy_2d^2p_{1T}d^2p_{2T}}&=&
 \sigma_0\left[x_af_A^\gamma(x_a)x_bf_B^\gamma(x_b)\right]_{b_\perp}\nonumber\\ &&\times\delta^{(2)}(\vec{p}_{1T}+\vec{p}_{2T}) \ ,
\end{eqnarray}
where $x_a=k_1/P_A$, $x_b=k_2/P_B$, $Q$ is the invariant mass for the produced lepton pair, and $Q^2=M_{\ell\ell}^2 = x_ax_b s$. $x_{a,b}$ represent the momentum fractions of incoming nucleon carried by the two incoming photons, and $s=(P_A+P_B)^2$ is the total hadronic center of mass energy squared per incoming nucleon pair. The leading order cross section $\sigma_0$ is defined as
\begin{equation}
    \sigma_0=\frac{|\overline {\cal M}_0|^2}{16\pi^2 Q^4}\ ,\label{e4}
\end{equation}
with the leading order amplitude squared 
\begin{equation}
|\overline {\cal M}_0|^2=(4\pi)^2\alpha_e^2\frac{2(t^2+u^2)}{tu} \ ,
\end{equation}
where $t$ and $u$ are usual Mandelstam variables for the $2\to 2$ process. To simplify the above expression, we have introduced an impact parameter $b_\perp$ dependent photon flux, which is also known as the joint photon distribution function,
\begin{eqnarray}
\left[f_A^\gamma(x_a)f_B^\gamma(x_b)\right]_{b_\perp}
&=&\int d^2b_{1\perp}d^2b_{2\perp}\delta^{(2)}(b_{1\perp}-b_{2\perp}-b_\perp)\nonumber\\
&&\times f_A^\gamma(x_a;b_{1\perp})f_B^\gamma(x_b;b_{2\perp}) \ ,\label{eq5}
\end{eqnarray}
where $f_{A,B}^\gamma(x_i;b_{i\perp})$ are individual photon distributions also referred as photon flux in the following discussion. These distributions describe the photon distribution at the transverse position $b_{i\perp}$ with respect to the center of the colliding nucleus. The above factorization can be regarded as a semi-classic picture of heavy ion collisions, where the impact parameter $b_\perp$ is related to the centrality of the collisions. For UPC events, $b_\perp$ is normally larger than $2R_A$ where $R_A$ is the nucleus radius. 

Incoming photons also carry non-zero transverse momenta, which has to be included because we are interested in the region with low transverse momentum imbalance for the lepton pair in the final state. In the leading order picture, the total transverse momentum imbalance of the lepton pair equals to the total transverse momentum of two incoming photons. In order to understand the final state interaction effects of the lepton pair with the hot medium in heavy ion collisions, such as the medium $p_T$-broadening, we need to have a precise descriptions for the initial state (of the photons) contributions. In the following, we will investigate these contributions.

\subsection{Impact Parameter Dependent Photon Fluxes}

We start with the analysis of the impact parameter dependent photon flux. In the last few years, there have been much interest in two-photon processes and photo-production processes in peripheral (non-UPC) heavy ion collisions~\cite{Klein:2018cjh,Zha:2018ywo,Aaboud:2018eph,Adam:2018tdm,Adam:2015gba}. For these events, we need to understand the photon flux beyond the simple all impact-parameter picture. The impact parameter dependent photon flux can be written as, see, for example Ref.~\cite{Vidovic:1992ik}
\begin{eqnarray}
xf_A^\gamma(x;b_\perp)=4Z^2\alpha 
\left|\int\frac{d^2{k_T}}{(2\pi)^2}e^{ik_T\cdot b_\perp}\frac{\vec{k}_T}{k^2}F_A(k^2)\right|^2 \ ,\label{fluxb1}  
\end{eqnarray}
where $k^2=k_T^2+x^2m_p^2$, $m_p$ is the nucleon mass, and $F_A(k^2)$ represents the normalized elastic charge form factor for the nucleus. For point-like particles ($F_A\equiv 1$), the photon flux becomes
\begin{equation}
xf^\gamma(x;b_\perp)=\frac{Z^2\alpha}{\pi^2} x^2m_p^2K_1^2(xm_pb_\perp) \ ,
\end{equation}
which is the well-known result in classical electrodynamics. The above formula has been applied to understand the dilepton production in non-UPC events in heavy ion collisions~\cite{Zha:2018ywo}, where the contributions from small impact parameter ($b_\perp <R_A$) played an important role. 

\begin{figure}[tbp]
\begin{center}
\includegraphics[width=7cm]{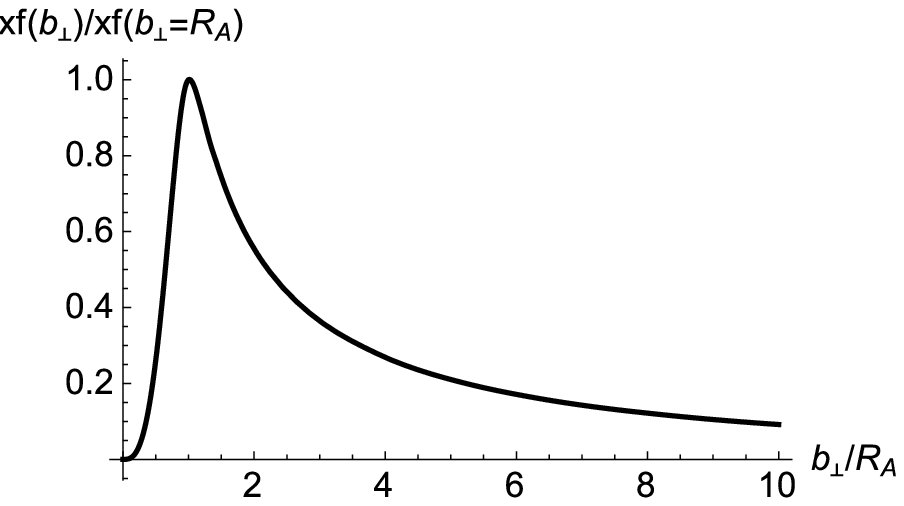}
\includegraphics[width=7cm]{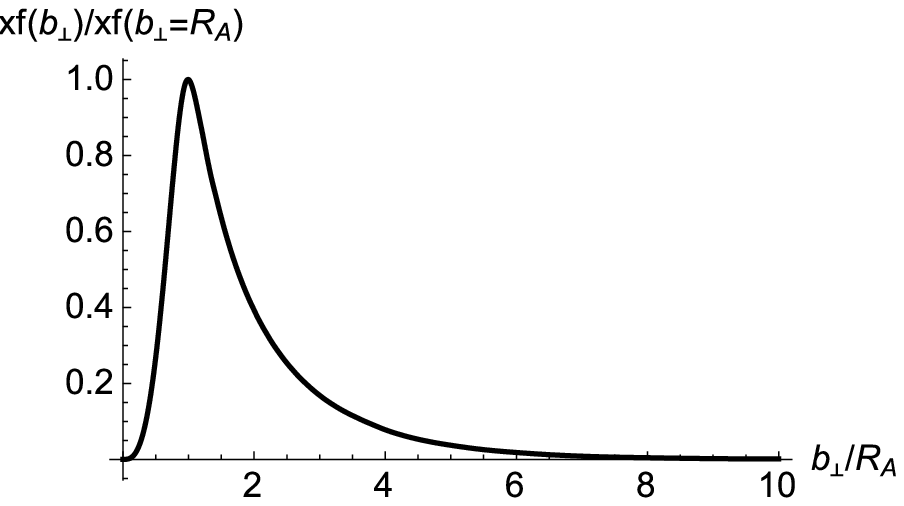}
\end{center}
\caption[*]{Impact parameter dependent photon flux as normalized to the flux at $b_\perp=R_A$ for a typical kinematics at the LHC with $x=10^{-3}$ and $R_A=7\rm fm$ (upper plot) and at RHIC with $x=10^{-2}$ (lower plot), respectively. }
\label{fluxb}
\end{figure}

There are two important features in this impact-parameter dependent photon flux. First, at large $b_\perp\gg R_A$, it reduced to the well-known Jackson result. This is because large $b_\perp$ leads to $k\to 0$ and thence to $F_A \to 1$.
This can be seen from Fig.~\ref{fluxb},  
where we plot the photon flux as function of $b_\perp$ for the typical kinematics at the LHC and RHIC. For the form factor, we follow the STARLIGHT~\cite{Klein:2016yzr},
\begin{eqnarray}
F_A(|\vec k|)&=&\frac{4\pi \rho^0}{|\vec k|^3\ A}\ \frac{1}{a^2 \vec k^2+1}\nonumber\\
&&\times \left [ \sin(|\vec k|R_A)-|\vec k|R_A \cos(|\vec k|R_A)\right ]\ , \label{ff}
\end{eqnarray}
where $R_A=7$~fm for $\rm Pb$, and $a=0.7$\ fm. 

Second, as shown in Fig.~\ref{fluxb}, at small-$b_\perp$, it is proportional to $b_\perp^3$. Here, the photon flux inside the nucleus is generated by the effective total charge of the nucleons inside the area denoted by $b_\perp$. 

\subsection{Transverse Momentum Dependence in the Photon Fluxes}

Except for Ref.~\cite{Hencken:2004td}, most previous studies ignored the inter-dependence between the impact parameter $b_\perp$ and the photon's transverse momentum. A recent attempt to address this issue was Ref.~\cite{Zha:2018tlq}, which extended the derivation of total cross section for two photon process in Ref.~\cite{Vidovic:1992ik} to differential cross section relevant to the STAR and ATLAS dilepton measurements. In the revised version of Ref.~\cite{Zha:2018tlq} and a recent paper by Li et al.~\cite{Li:2019yzy, Li:2019sin}, the so-called QED approach~\cite{Hencken:2004td} has been applied to compute the dilepton production in two photon processes with full dependence on the impact parameter and the pair $p_T$.

Here, we investigate this from different point of view, following the factorization argument and studying the individual photon flux. This result may also be relevant when considering the photon $p_T$ contribution to the final state $p_T$ in photoproduction, especially imaging studies~\cite{Klein:2019qfb}. When we integrate out the impact parameter, the transverse momentum distribution can be evaluated as~\cite{Klein:1999gv}
\begin{equation}
xf_\gamma^A(x,k_T) = \frac{Z^2 \alpha}{\pi^2}\frac{k^2_T}{\left(k_T^2+x^2m_p^2\right)^2}F_A^2(k^2)\ , \label{nucleus}
\end{equation}
where again $k^2=\left(k_T^2+x^2m_p^2\right)$ and $F_A$ is the nuclear charge form factor. This has been widely employed to estimate the transverse momentum dependence in two photon processes in UPCs, see, e.g., the STARLIGHT simulation~\cite{Klein:2016yzr}.

It is non-trivial to derive the impact-parameter and transverse momentum dependent photon flux. The main difficulty is that the impact parameter $b_\perp$ is Fourier conjugate variable associated with the photon's transverse momentum $k_T$. There will be model dependence to compute the combined distribution from the classic EM fields. In the following, we will estimate the average transverse momentum squared, and comment on the difficulty to calculate the combined distribution directly.

Before we get to the details of the models, we would like to emphasize some important features on the transverse momentum distribution for the incoming photons in nucleus. First, for $b_\perp\ll R_A$, effective charge contribution is limited to protons inside $b_\perp$ region, and the flux is proportional to $1/b_\perp^2$ for the average transverse momentum squared. This is an important feature which should be satisfied by all model calculations. Similarly, at large $b_\perp\gg R_A$, because of uncertainty principle, the average squared transverse momentum squared is also of order $1/b_\perp^2$. Around $b_\perp\sim R_A$, on the other hand, the average transverse momentum may differ from the above parametric estimates. However, with the constraints from these generic features ($b_\perp\ll R_A$ and $b_\perp\gg R_A$), the model calculations of the average transverse momentum squared are very much determined. 

In Eq.~(\ref{fluxb1}) the integral variable $k_T$ is the photon's transverse momentum. We may compute the average transverse momentum squared by multiplying the integrand by the products of $k_T$, 
\begin{eqnarray}
\langle k_T^2\rangle&=&\frac{4Z^2\alpha}{xf^\gamma(x;b_\perp)}\int\frac{d^2{k_T}}{(2\pi)^2}\frac{d^2{k_T'}}{(2\pi)^2}e^{i(k_T-k_T')\cdot b_\perp}\nonumber\\
&& ~~~~~~\times\frac{({k}_T\cdot {k}_T')^2}{k^2k^{\prime 2}}F_A(k^2)F_A(k^{\prime 2})\label{e10}  \ , \end{eqnarray}
where we manipulate the square of the integral in Eq.~(\ref{fluxb1}) with two separate integrals. The average transverse momentum squared is computed by weighting the integrals with the scalar product of $k_T$ and $k_T'$. Certainly, there is model dependence on how we weight the integrals. The above formula reproduces the average transverse momentum squared for minimum bias case, which provides a cross check for the above estimate. 

We can further simplify the above equation by integrating out the azimuthal angles. For example, Eq.~(\ref{fluxb1}) can be written as,
\begin{eqnarray}
xf^\gamma(x;b_\perp)=4Z^2\alpha \frac{|c_1(x;b_\perp)|^2}{b_\perp^2} \ ,
\end{eqnarray}
where $c_1(x;b_\perp)$ is defined as
\begin{eqnarray}
c_1(x;b_\perp)=\int dk_T^2 \frac{b_\perp k_T J_1(k_T b_\perp)}{k^2}F_A(k^2) \ ,
\end{eqnarray}
with $J_1(x)$ the $J$-type generalized Bessel function. 

\begin{figure}[tbp]
\begin{center}
\includegraphics[width=7cm]{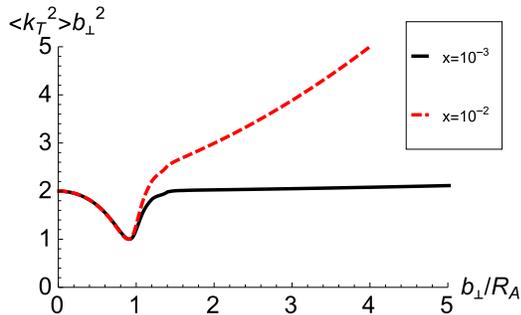}
\end{center}
\caption[*]{Average transverse momentum squared multiplied by $b_\perp^2$ for the photon distributions as function of the impact parameter $b_\perp$ at $x=10^{-3}$ and $x=10^{-2}$. Clearly, both cases predict a generic behavior of $\langle k_T^2\rangle \sim 1/b_\perp^2$, in particular for small $b_\perp$. At large $b_\perp$ for $x=10^{-2}$, the above relation breaks down because the average transverse momentum is now comparable to $xm_p$ with $x=10^{-2}$.}
\label{avekt2}
\end{figure}

Similarly, we can work out the integrals over the azimuthal angles in Eq.~(\ref{e10}). With that, we arrive at a very simple result for the average transverse momentum squared,
\begin{eqnarray}
\langle k_T^2\rangle (x;b_\perp)&=&\frac{1}{b_\perp^2}\left[1+\left(1-\frac{c_2(x;b_\perp)}{c_1(x;b_\perp)}\right)^2\right] \ ,
\end{eqnarray}
where $c_2(x;b_\perp)$ is defined as
\begin{eqnarray}
c_2(x;b_\perp)=\int dk_T^2 \frac{b_\perp^2 k_T^2 J_2(k_T b_\perp)}{k^2}F_A(k^2) \ .
\end{eqnarray}
The above results have a very nice feature: the explicit dependence of average transverse momentum squared $\langle k_T^2\rangle$ on the impact parameter $b_\perp^2$. The overall behavior $\langle k_T^2\rangle \propto 1/b_\perp^2$ is consistent with the above generic discussions. In addition, we also find that the additional factor only has a mild dependence on $b_\perp$. As an example, in Fig.~\ref{avekt2}, we show the typical case for the LHC at $x=10^{-3}$ and RHIC at $x=10^{-2}$. For the latter case, because $xm_p$ is comparable to $1/b_\perp$ at large $b_\perp$, the above relation will be modified accordingly.

\subsection{Photon-photon Interaction Rate in Heavy Ion Collisions}

\begin{figure}[tbp]
\begin{center}
\includegraphics[width=7cm]{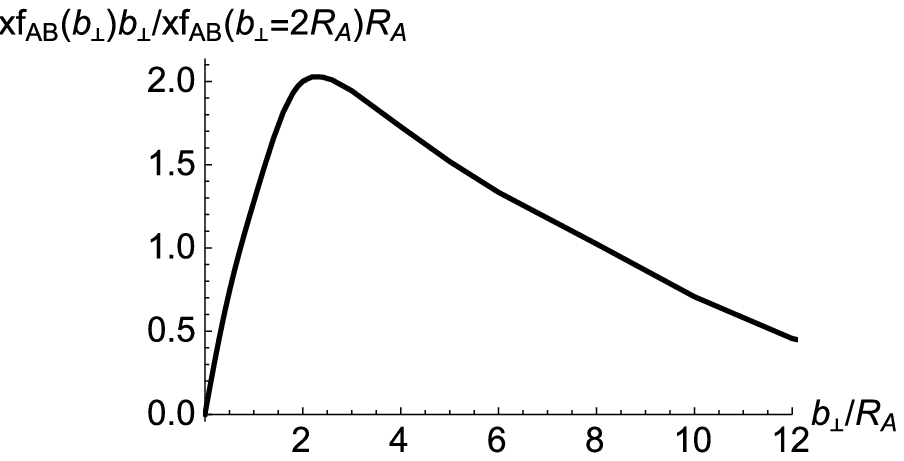}
\includegraphics[width=7cm]{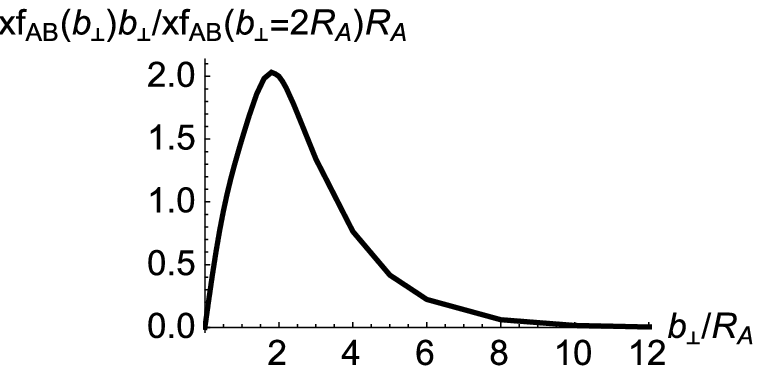}
\end{center}
\caption[*]{Relative collision rate in collisions between nuclei $A$ and $B$ as function of the collision impact parameter $b_\perp$ assuming $x_a=x_b=10^{-3}$ for the typical kinematics at the LHC (upper plot) and $x_a=x_b=10^{-2}$ for RHIC (lower plot). We have normalized the distributions to that at $b_\perp=2R_A$.}
\label{fluxAB}
\end{figure}

With the photon flux for each of the incoming nucleus, we can calculate the two-photon interaction rate following Eq.~(\ref{eq5}),
\begin{eqnarray}
x_ax_bf_{AB}(b_\perp)&=&\int d^2b_{1\perp}d^2b_{2\perp}\delta^{(2)}(b_\perp-b_{1\perp}-b_{2\perp})\nonumber\\
&&\times x_af_A(b_{1\perp})x_bf_B(b_{2\perp})\ ,\label{fluxABe}
\end{eqnarray}
where $f_{A,B}(b_\perp)$ represents the individual photon fluxes for $A$ and $B$ nucleus, following the definition in Eq.~(\ref{fluxb1}) and
$x_{a,\, b}$ are the photon longitudinal momentum fractions. Figure~\ref{fluxAB} shows the relative interaction rate as function of the impact parameter for the typical kinematics at the LHC and RHIC. The relative interaction rate increases with impact parameter and peaks around $b_\perp\sim 3R_A$ region before decreasing as the impact parameter increases further. In minimum bias samples, most of the dileptons are produced in UPCs.

\begin{figure}[tbp]
\begin{center}
\includegraphics[width=7cm]{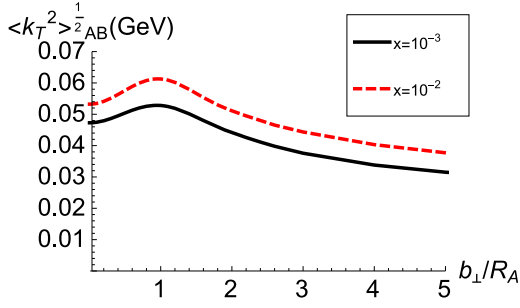}
\end{center}
\caption[*]{Average transverse momentum squared for the two-photon final state in ion-$A$ on ion-$B$ collisions as function of the impact parameter $b_\perp$ for heavy ion collisions.}
\label{avektAB}
\end{figure}

We can also estimate the average transverse momentum squared for the collisions, by adding the contributions from both incoming photons' average transverse momentum squared. 
Figure ~\ref{avektAB} shows the total transverse momentum squared as function of $b_\perp$ for $x_a=x_b=10^{-3}$ and $x_a=x_b=10^{-2}$, respectively. We see that the average transverse momentum does not change dramatically depending on the impact parameter. In particular, from peripheral to central collisions, it remains nearly constant. It seems that the average transverse momentum squared contribution from the incoming photons depends mildly on the collision centrality. Comparing to the results from Ref.~\cite{Zha:2018tlq}, we find that our above results predict quite different behavior for the average transverse momentum squared in the two photon processes in heavy ion collisions. This may be due to different treatments of impact parameter dependences. Therefore, in the following, we try to generalize the EPA approach by introducing the photon Wigner distribution, which contains both the impact parameter and transverse momentum information simultaneously.

\subsection{Generalized Equivalent Photon Approximation: Photon Wigner Distribution}
In order to study the joint transverse momentum and impact parameter dependence in the photon flux, we follow the parton Wigner distribution functions introduced in Ref.~\cite{Belitsky:2003nz} to introduce the photon Wigner distribution, 
\begin{eqnarray}
   && xf_\gamma(x,k_T ;b_\perp) \notag \\
   &&=\int\frac{d^2\Delta_\perp}{(2\pi)^2}e^{i\Delta_\perp\cdot b_\perp}\int\frac{d\xi^-d^2r_\perp}{(2\pi)^3}e^{ixP^+\xi^--ik_T\cdot r_\perp}\nonumber\\  
    && \langle A,-\frac{\Delta_\perp}{2}|F^{+\perp}\left(0,\frac{r_\perp}{2}\right)F^{+\perp}\left(\xi^-,-\frac{r_\perp}{2}\right)|A,\frac{\Delta_\perp}{2}\rangle ,\, \,
\end{eqnarray}
where a momentum difference $\Delta_\perp$ has been introduced for the nucleus states to obtain the impact parameter dependence. 
For convenience and simplicity, we also introduce the so-called generalized TMD (GTMD) photon distributions in the momentum space,
\begin{eqnarray}
   && \Gamma^{ij}(x,k_T;\Delta_\perp)=\int\frac{d\xi^-d^2r_\perp}{(2\pi)^3}e^{ixP^+\xi^--ik_T\cdot r_\perp}\\
    &&\times \langle A,-\frac{\Delta_\perp}{2}|F^{+i}\left(0,\frac{r_\perp}{2}\right)F^{+j}\left(\xi^-,-\frac{r_\perp}{2}\right)|A,\frac{\Delta_\perp}{2}\rangle \ , \nonumber
\end{eqnarray}
where we keep the transverse indices open to construct various GTMDs. For our purpose, we deduce the following parameterization,
\begin{eqnarray}
\Gamma^{ij}(x,k_T;\Delta_\perp)&&=\frac{\delta^{ij}}{2}xf_\gamma(x,k_T;\Delta_\perp) \nonumber \\
&&+\left(\frac{k_+^ik_-^j}{\vec k_+\cdot \vec k_-}-\frac{\delta^{ij}}{2}\right) xh_\gamma(x,k_T;\Delta_\perp), \,\,\, 
\end{eqnarray}
where $k_\pm=k_T\pm \Delta_\perp/2$, $f_\gamma$ represents the usual photon distribution and $h_\gamma$ stands for the so-called linearly-polarized photon distribution. For the convenience of the following derivations, we choose a particular decomposition of the above linearly-polarized GTMD $h_\gamma$. We emphasize that different parameterization can be applied. 
When we integrate the Wigner distribution over the impact parameter $b_\perp$, the above reproduces the TMD photon distributions introduced in Ref.~\cite{Li:2019yzy,Li:2019sin}. 

In the classical limit and analogous to the QCD case\cite{Hatta:2016dxp}, we can find that the above GTMD photon distribution for a heavy nucleus with charge $Ze$ can be written as
\begin{eqnarray}
xf_\gamma(x,k_T;\Delta_\perp)&=&xh_\gamma(x,k_T;\Delta_\perp)\nonumber\\
&=&\frac{4Z^2\alpha}{(2\pi)^2}  \frac{q_\perp\cdot q_\perp'}{q^2 q^{\prime 2}} F_A(q^2)F_A(q'^2) \ ,
\end{eqnarray}
where $F_A$ is nucleus form factor, $q_\perp=k_T-\Delta_\perp/2$, $q_\perp'=k_T+\Delta_\perp/2$ and $q^2=q_\perp^2+x^2m_p^2$.

\begin{figure}[tbp]
\begin{center}
\includegraphics[width=7cm]{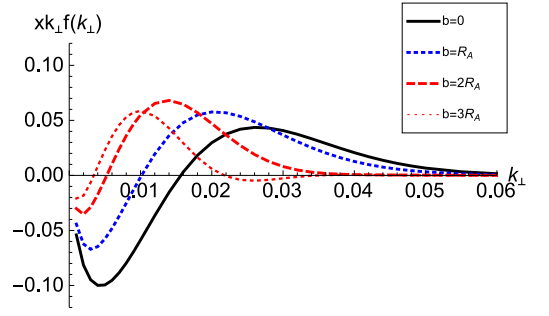}
\end{center}
\caption[*]{$k_\perp$ distribution calculated from Eq.~(\ref{wigner}) for different impact parameters with arbitrary unit.}
\label{noncentralkt}
\end{figure}

As an example, we can obtain the Wigner distribution for the usual photon flux as the Fourier transform of the above GTMD,
\begin{eqnarray}
    xf_\gamma(x,k_T;b_\perp)&=& \int\frac{d^2\Delta_\perp}{(2\pi)^2}e^{i\Delta_\perp\cdot b_\perp} xf_\gamma(x,k_T;\Delta_\perp) . \label{wigner}
\end{eqnarray}
As expected from the property of the Wigner distribution, it is straightforward to check that Eq.~(\ref{wigner}) reduces to the $b_\perp$ distribution in Eq.~(\ref{fluxb1}) and the $k_\perp$ distribution in Eq.~(\ref{nucleus}) after integrating over $k_\perp$ and $b_\perp$, respectively. However, the above formulas do not have a clear probabilitistic interpretation as the impact parameter and transverse momentum dependent photon distributions in the factorized cross section calculations. For example, the above formula predict an oscillation function of $k_\perp$ depending on different values of $b_\perp$, shown in Fig.~\ref{noncentralkt}. That means the distribution function calculated from the above equations is not guaranteed to be positive definite. However, when one apply the photon Wigner distribution or the photon GTMD in the two photon fusion processes, one can demonstrate that the corresponding factorized cross section is positive definite, which will be discussed in the following subsection.

\subsection{Generalized Equivalent Photon Approximation}

In order to take into account the transverse momentum dependence for the incoming photon fluxes in heavy ion collisions, we need to generalize the equivalent photon approximation (EPA). In previous EPA models, such as that implemented in STARLIGHT, the transverse momentum dependence was introduced as an average over all impact parameter. In the following, we extend this approximation by applying the photon Wigner distribution introduced above, which we refer to the generalized EPA (GEPA).

In our consideration, two nuclei collide at a particular impact parameter $b_\perp$, which represents the transverse distance between the centers of the two nuclei. To produce the lepton pair through short distance process of $\gamma+\gamma\to \ell^+\ell^-$, we assume the interaction point is at the transverse place with $b_{1\perp}$ and $b_{2\perp}$ respect to the nuclei centers. In this configuration, the individual photon flux will depend on $b_{1\perp}$ and $b_{2\perp}$, respectively. Clearly, the collision impact parameter can be determined by $\vec{b}_\perp=\vec{b}_{1\perp}-\vec{b}_{2\perp}$. Therefore, we can write down a generic factorization formula for dilepton production in $AA$ collisions through $\gamma+\gamma\to \ell^+\ell^-$ as,
\begin{eqnarray}
&&\frac{d\sigma(AB[\gamma\gamma]\to\mu^+\mu^-)}{dy_1dy_2d^2p_{1\perp}d^2p_{2\perp}}=
\int d^2b_{1\perp}d^2b_{2\perp}\delta^{(2)}(b_\perp-b_{1\perp}+b_{2\perp})\nonumber\\
&&~~~~~~~~~\times \Gamma^{ij}(k_{1T},b_{1\perp})\Gamma^{kl}(k_{2T},b_{2\perp})\nonumber\\
&&~~~~~~~~~\times H_{ijkl}(P_\perp) \delta^{(2)}(p_T-k_{1T}-k_{2T})\ , 
\end{eqnarray}
where $ik$ represent the polarization indices for incoming two photons in the scattering amplitude and $jl$ for the complex conjugate of the amplitude, $\Gamma^{ij}$ and $\Gamma^{kl}$ stand for the two photon Wigner distributions. Again, we will focus on the kinematic region with correlation limit where the total transverse momentum of the lepton pair $\vec{p}_\perp$ is much smaller than the individual transverse momentum $p_{1\perp}$ or $p_{2\perp}$. Therefore, we can factorize the cross section into hard part $H_{ijkl}$ and soft part $\Gamma^{ij}$ and $\Gamma^{kl}$. In partiuclar, the hard part $H_{ijkl}$ only depends on the hard momentum scale $P_\perp$. In this paper, we focus on the azimuthal angular average cross sections for individual lepton, i.e., integrating over the azimuthal angle of the lepton ($\vec{p}_{1\perp}$ or $\vec{p}_{2\perp}$). In the end, one can show that the hard part can be written as
\begin{eqnarray}
H_{ijkl}=\sigma_0\left[\delta^{ij}\delta^{kl}-\delta^{ik}\delta^{jl}+\delta^{il}\delta^{jk}\right] \ ,
\end{eqnarray}
where $\sigma_0 =\frac{1}{\pi} \frac{d\sigma}{dt} =\frac{2\alpha^2}{\hat{s}^2}\left(\frac{\hat u}{\hat t} +\frac{\hat t}{\hat u}\right)$. Substituting the Wigner distribution parameterizations in Sec.~IIC, we will obtain the following expression for the differential cross section,
\begin{eqnarray}
\frac{d\sigma}{d\Omega}&=&
\sigma_0\int d^2\Delta_{\perp}d^2k_{1\perp}d^2k_{2\perp}\frac{e^{i\Delta_\perp\cdot b_\perp}}{(2\pi)^2}
\delta^{(2)}(p_\perp-k_{1\perp}-k_{2\perp})\nonumber\\
&&\times\left[x_1f_\gamma(x_1,k_{1\perp};\Delta_\perp)x_2f_\gamma(x_2,k_{2\perp};\Delta_\perp)+\right.\nonumber\\
&&\left. + x_1h_\gamma(x_1,k_{1\perp};\Delta_\perp)x_2h_\gamma(x_2,k_{2\perp};\Delta_\perp){\cal H}_{hh}\right]\ , 
\end{eqnarray}
where $d\Omega=dy_1dy_2d^2p_{1\perp}d^2p_{2\perp} d^2 b_\perp$ and ${\cal H}_{hh}$ stands for
\begin{eqnarray}
{\cal H}_{hh}=\frac{\vec k_{1+}\cdot \vec k_{2-}\vec k_{1-}\cdot \vec k_{2+}-\vec k_{1+}\cdot \vec k_{2+}\vec k_{1-}\cdot \vec k_{2-}}{\vec k_{1+}\cdot \vec k_{1-}\vec k_{2+}\cdot \vec k_{2-}} \ .
\end{eqnarray}

\begin{widetext}
With the results for $f_\gamma$ and $h_\gamma$ in previous sub-section, the above reproduces the results in Ref.~\cite{Li:2019sin}. As also noted in Ref.~\cite{Li:2019sin}, the second term in the differential cross section vanishes when we integrate over either $b_\perp$ or $p_\perp$. Therefore, for a particular impact parameter $b_\perp$, it leads to an oscillation contribution as function of $p_\perp$. In terms of incoming photon momenta, after averaging over the azimuthal orientation of the lepton pair, the cross section for the production of lepton pair in AA collisions at fixed impact parameter $b_\perp$ can be cast into
\begin{eqnarray}
\frac{d \sigma}{dy_1 dy_2 d^2 p_{1T} d^2 p_{2T} d^2 b_\perp} &= &\int d^2 k_{1\perp} d^2 k_{1\perp}^\prime \frac{1}{(2\pi)^2}e^{i(k_{1\perp}-k_{1\perp}^\prime )\cdot b_\perp} \left(\frac{Z^2 \alpha}{\pi}\right)^2 \frac{F(k_1^2)}{k^2_1}\frac{F(k_1^{\prime 2})}{k^{\prime 2}_1}\frac{F(k_2^2)}{k^2_2}\frac{F(k_2^{\prime 2})}{k^{\prime 2}_2} \notag \\
&& \times \sigma_0  \left[(k_{1\perp} \cdot k_{1\perp}^\prime) (k_{2\perp} \cdot k_{2\perp}^\prime) -(k_{1\perp} \times k_{1\perp}^\prime)\cdot (k_{2\perp} \times k_{2\perp}^\prime)\right], \label{fact}
\end{eqnarray}
where $k_i^2 = x_i ^2 M^2 +k_{i\perp}^2$, $\vec{k}_{2\perp} = \vec{p}_T - \vec{k}_{1\perp}$ and $\vec{k}^\prime_{2\perp} = \vec{p}_T - \vec{k}^\prime_{1\perp}$. It is straightforward to check that the contribution from the linearly polarized photons ${\cal H}_{hh}$, i.e., the second term inside the square brackets in Eq.~(\ref{fact}) vanishes when one integrates over the impact parameter $b_\perp$. Thus, we can reproduce the previous known integrated cross section for lepton pair productions. Usually the GTMD or its corresponding Wigner distribution is not positive definite. It is quite interesting to note that the cross section is in fact positive definite, since it can be written as 
\begin{eqnarray}
\frac{d \sigma}{dy_1 dy_2 d^2 p_{1T} d^2 p_{2T} d^2 b_\perp} &= &  \sigma_0 G^{ik} G^{jl \ast} \left[ \delta^{ij} \delta^{kl} - \delta^{ik}\delta^{jl} + \delta^{il}\delta^{jk}\right] \notag \\
&=& \sigma_0\left[ (G^{11} -G^{22}) (G^{11\ast} -G^{22 \ast})+ (G^{12} +G^{21}) (G^{12\ast} +G^{21 \ast}) \right] 
\end{eqnarray}
where $G^{ik} =\int \frac{d^2 k_{1\perp} }{(2\pi)}e^{ik_{1\perp}\cdot b_\perp} k_{1\perp}^i k_{2\perp}^k  \frac{F(k_1^2)}{k^2_1}  \frac{F(k_2^2)}{k^2_2}$. It is also worth mentioning that the above cross section vanishes when $b_\perp =0$ and $p_T =0$. This can be shown by noting that $G^{11}$ becomes the same as $G^{22}$ and $G^{12} =- G^{21}$ when one sets $b_\perp =0$ and $p_T =0$. This essentially explains the existence of the so-called displaced peaks in the measurement of the momentum imbalance. 
\end{widetext}

\subsection{Numerical results in the GEPA Framework}

\begin{figure}[h]
 \includegraphics[height=63mm]{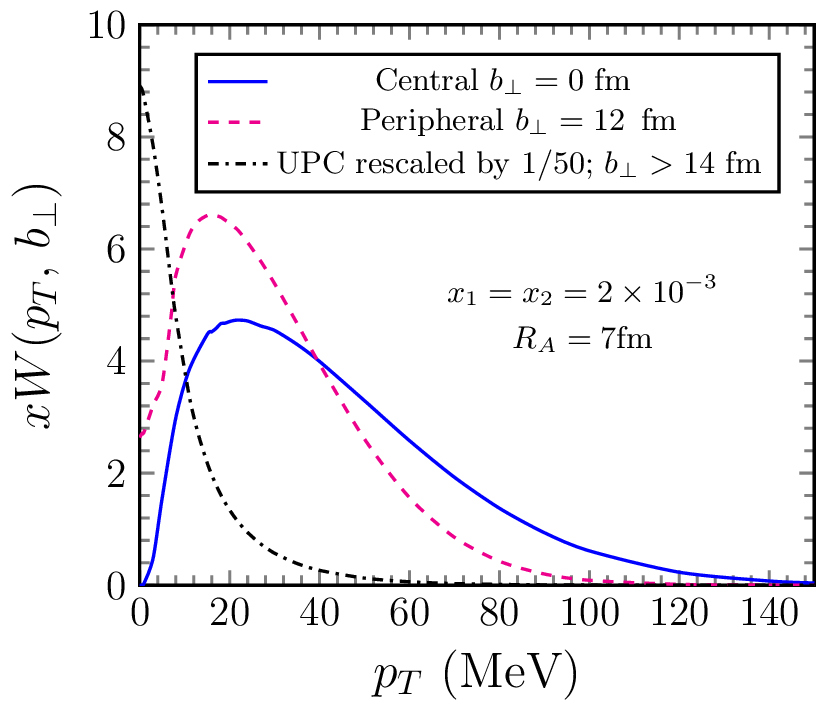} 
\caption{Transverse momentum imbalance distribution $xW (x, p_T, b_\perp)$ with different values of impact parameters $b_\perp$. The Gaussian form factor $F(k^2) =\exp [-k^2/Q_0^2]$ with $Q_0 = 80 \text{MeV}$ and the Wood-Saxon type of form factor yield almost identical numerical results.  
} 
\label{fig2}
\end{figure}

\begin{figure}[h]
 \includegraphics[height=60mm]{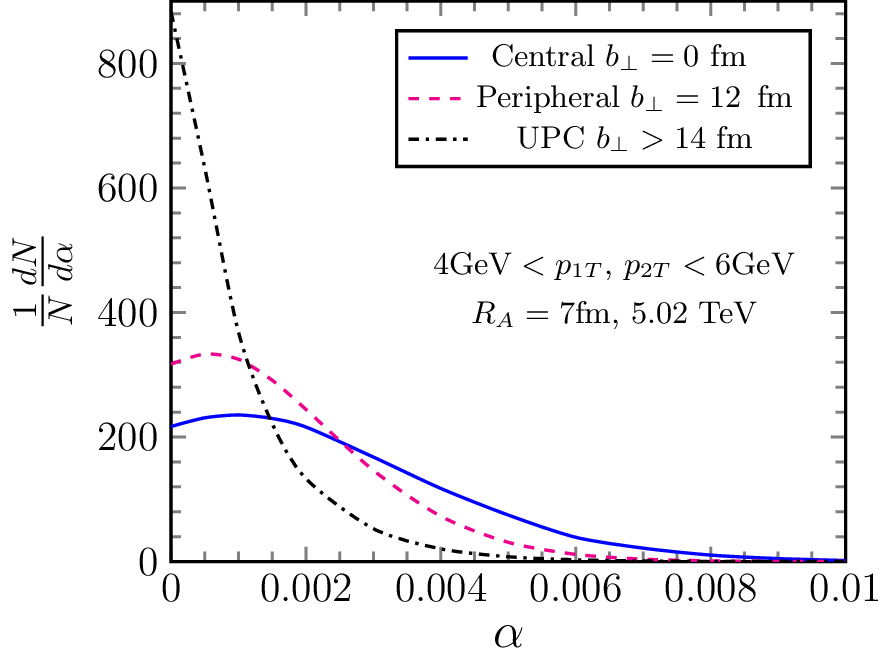}
\caption{The resulting normalized $\alpha \equiv 1-\frac{|\phi_1 -\phi_2 |}{\pi}$ distribution.} 
\label{alphafig}
\end{figure}

To perform the numerical evaluation, let us first average over the azimuthal angle of the impact parameter $b_\perp$, then the corresponding formula reads
\begin{eqnarray}
\frac{d \sigma}{  \pi d b^2_\perp d \Delta \phi} & =& (2\pi) \int dy_1 dy_2  \int dp_{1\perp} p_{1\perp}  \int dp_{2\perp} p_{2\perp} \notag \\
&& \times  xW (x, p_T, b_\perp) \sigma_0
\end{eqnarray}
where $ xW (x, p_T, b_\perp)$, which encodes the initial momentum imbalance, is defined as
\begin{eqnarray}
 xW (x, p_T, b_\perp)  &=& \int d^2 k_{1\perp} d^2 k_{1\perp}^\prime \frac{ J_0(|k_{1\perp}-k_{1\perp}^\prime| b_\perp )}{(2\pi)^2} \notag \\
 && \times x_1 f^{ij}(x_1, k_{1\perp}, k_{1\perp}^\prime) x_2 f^{kl}(x_2, k_{2\perp}, k_{2\perp}^\prime) \notag \\
 &&\times    \left[ \delta^{ij} \delta^{kl} - \delta^{ik}\delta^{jl} + \delta^{il}\delta^{jk}\right] ,
\end{eqnarray}
where $\vec{p}_T = \vec{p}_{1\perp} +\vec{p}_{2\perp}$. Furthermore, for UPC, we can integrate over $b_\perp$ from $2R_A$ to $\infty$, and find
\begin{eqnarray}
\frac{d \sigma_{\text{UPC}}}{dy_1 dy_2 d^2 p_{1T} d^2 p_{2T} } &= & (4\pi R_A^2)   xW_{\text{UPC}} (x, p_T, b_\perp) \, \sigma_0, \,\,\,
\end{eqnarray}
where $xW_{\text{UPC}} (x, p_T, R_A) $ can be similarly written as
\begin{eqnarray}
&& xW_{\text{UPC}} (x, p_T, R_A)  =\frac{1}{4\pi R_A^2} \int d^2 k_{1\perp} d^2 k_{1\perp}^\prime  \notag \\
&& \quad  \times \left[ \delta^{(2)}(k_{1\perp} - k_{1\perp}^\prime)-\frac{ 2R_A \, J_1(2R_A |k_{1\perp}-k_{1\perp}^\prime| )}{(2\pi) |k_{1\perp}-k_{1\perp}^\prime|}  \right]\notag \\
 && \quad \times  x_1 f^{ij}(x_1, k_{1\perp}, k_{1\perp}^\prime) x_2 f^{kl}(x_2, k_{2\perp}, k_{2\perp}^\prime) \notag \\
 && \quad \times  \left[ \delta^{ij} \delta^{kl} - \delta^{ik}\delta^{jl} + \delta^{il}\delta^{jk}\right]. 
\end{eqnarray}
Here we normalize $xW_{\text{UPC}} (x, p_T, R_A) $ by a factor of $\frac{1}{4\pi R_A^2}$ to make it dimensionless. 

The numerical evaluation of the above results are shown in Figs.~\ref{fig2} and \ref{alphafig}. As anticipated, there is a dip or a displaced peak in the distribution of the momentum imbalance $p_T$ in the low $p_T$ region for the central collisions, since the $xW(x, p_T, b_\perp)$ function vanishes at $b_\perp =0$ and $p_T =0$. In terms of the normalized $\alpha$ distribution measured first by the ATLAS experiment as shown in Fig.~\ref{alphafig}, the displaced peak becomes less prominent due to the smearing effect after averaging over the individual momenta of the produced leptons. We have also checked that the GEPA calculation also agrees with the ATLAS data measured in the peripheral collisions. This indicates that the transverse momentum broadening effect measured in the peripheral collisions can be captured by the incoming coherent photon distributions. In contrast, the transverse momentum broadening due to the coherent incoming photons is not sufficient to accurately describe the data in the central collisions, where additional broadening effects due to the incoherent multiple scattering and other medium effects as well as Sudakov type soft photon emissions may become important. We believe that further theoretical developments as well as additional measurements\cite{CMS:2020avp} and experimental efforts including the possible unfolding and correction of the data due to detector responses can help us reach more conclusive phenomenological findings. \\

\section{Soft Photon Radiation and Sudakov Resummation}

Higher order QED corrections as shown in Fig.~\ref{fig1}(b,c) can modify the leading order picture described in previous subsection. In the following, we focus on soft photon radiation, which have strong effects on the $p_T$ distributions for the lepton pair especially at moderately large acoplanarity.  

At one-loop order, soft photon radiations from the final state leptons dominate the small transverse momentum for the pair. The soft photon radiation from the lepton propagator is power suppressed for large transverse momentum lepton production by order of $k_{s\perp}/P_T$, where $k_{s\perp}$ is the radiated photon's transverse momentum and $P_T$ for the individual lepton's transverse momentum. Therefore, this type of diagram is discarded in the leading power approximation. 

We apply the Eikonal approximation to calculate the soft photon radiation contribution, see, e.g.~\cite{Sun:2015doa},
\begin{equation}
{\cal M}^{(1)r}|_{\rm soft}=e \left(\frac{p_1^\mu}{p_1\cdot k_s}-\frac{p_2^\mu}{p_2\cdot k_s}\right){\cal M}^{(0)} \ ,
\end{equation}
where ${\cal M}^{(0)}$ represents the leading order Born amplitude, $\mu$ is the polarization index for the soft photon with momentum $k_s$, and the minus sign in the bracket comes from the fact that the contributions from the lepton and anti-lepton differ by a minus sign for the soft photon radiation. Therefore, the contribution to the amplitude squared reads as 
\begin{equation}
|{\cal M}^{(1)r}|^2_{\rm soft}=e^2\frac{2p_1\cdot p_2}{p_1\cdot k_s p_2\cdot k_s}|{\cal M}^{(0)}|^2 \ .
\end{equation}
This additional soft photon radiation generates additional non-zero transverse momentum for the lepton pair, and the consequence is that the lepton pair will no longer be in the back-to-back direction in the transverse momentum plane. In ATLAS, the imbalance between the lepton and anti-lepton was measured through the azimuthal angular correlation between them. In order to study the azimuthal angular distribution of the lepton pair, we calculate the total transverse momentum generated by the radiated photon. Together with the incoming photons' contributions as discussed in previous section, this lead to the final total transverse momentum distribution for the lepton pair. Alternatively, we also find that we can carry out this calculation by studying the radiated photon contribution to the ``transverse momentum" imbalance in the lepton frame, which can be translated into azimuthal angular distribution between the two leptons in the Lab frame. As shown in Appendix \ref{frame}, these two frames are consistent in generating the azimuthal angular correlation between the two leptons in the final state.

We will show the derivation in the lepton frame, and comment on the Lab frame calculations later. In the lepton frame, the two leptons are moving back-to-back along $\hat z$-direction. Soft photon radiation give the lepton pair a small additional transverse momentum $\ell_\perp$. Now including the lepton mass (assumed to be $\mu$ mass here, for convenience), we have $p_1^2=p_2^2=m_\mu^2$ and the real diagram contribution to the soft photon radiation can be written as
\begin{eqnarray}
S^{(r)}(\ell_\perp)=\frac{\alpha_e}{\pi^2}\int\frac{d\xi}{\xi}\frac{\ell_\perp^2}{(\ell_\perp^2+\xi^2m_\mu^2)(\ell_\perp^2+\bar\xi^2m_\mu^2)} \ ,
\end{eqnarray}
in the transverse momentum space, where $\xi=k_s\cdot p_2/p_1\cdot p_2$ and $\bar\xi=k_s\cdot p_1/p_2\cdot p_1$. It is clear that lepton mass plays an important role here - the lighter the lepton, the more Sudakov radiation. 
From the kinematics, we know that $\xi\bar \xi=\ell_\perp^2/Q^2$. Furthermore, $\xi$ integral is limited by $\ell_\perp^2/Q^2<\xi<1$. Carrying out $\xi$ integration leads to the following expression for the soft photon radiation at small transverse momentum of $\ell_\perp$,
\begin{equation}
S^{(r)}(\ell_\perp)=\frac{\alpha}{\pi^2}\frac{1}{\ell_\perp^2}\ln\frac{Q^2}{\ell_\perp^2+m_\mu^2} \ .\label{soft0}
\end{equation}
When $\ell_\perp\gg m_\mu$, the above result leads to a double logarithmic behavior as $1/\ell_\perp^2\ln(Q^2/\ell_\perp^2)$, which is very much similar to the behavior of the back-to-back hadron production in $e^+e^-$ annihilation, $e^+e^-\to h_1h_2+X$, first studied in Refs.~\cite{Collins:1981uk,{Collins:1981uw},{Collins:1981va}}. This double logarithmic behavior can be factorized into two fragmentation functions depending on the transverse momentum $\ell_\perp$, and the relevant resummation can be carried out. However, in our current case, because of the lepton mass, the above distribution scales as $1/\ell_\perp^2\ln(Q^2/m_\mu^2)$ when $\ell_\perp\ll m_\mu$. This leads a totally different infrared behavior as $\ell_\perp\to 0$. In the sense, we only have soft divergence at this limit, which, of course, will be canceled out by the virtual diagrams. With that cancellation, we will be able to derive the complete result at one-loop order. In the following, we will focus on the large logarithms at this order because they dominate the differential cross section contributions. These large logarithms arise from the soft photon radiation in both real and virtual diagrams.

The logarithms become more evident when we Fourier transform the above real radiation contribution into $r_\perp$ space conjugate to $\ell_\perp$,
\begin{equation}
S_u^{(r)}(r_\perp)=\frac{\alpha}{\pi^2}\int\frac{d^2\ell_\perp}{\ell_\perp^2}e^{i\ell_\perp\cdot r_\perp}\ln\frac{Q^2}{\ell_\perp^2+m_\mu^2} \ .
\end{equation}
Compared to the di-hadron correlation in $e^+e^-$ annihilation studied in~\cite{Collins:1981va}, in our case, there are additional complexity because of the lepton mass. The muon mass $m_\mu\approx 0.1\rm GeV$ is relevant because of the total transverse momentum for the lepton pair is in the same range as the lepton mass. Therefore, the large logarithms will depend on the relative size between $\mu_r$ ($\mu_r\approx 1/r_\perp$) and the lepton mass $m_\mu$. When $\mu_r>m_\mu$, it corresponds to the transverse momentum $\ell_\perp$ larger than the lepton mass, and we will have a similar double logarithmic term as that in $e^+e^-\to h_1h_2+X$ studied in Ref.~\cite{Collins:1981va}. This can be understood as the following cancellation between the real and virtual contributions,
\begin{eqnarray}
&&S_u^{(r)}(r_\perp)+S_u^{(v)}(r_\perp)|_{\mu_r>m_\mu}\nonumber\\
&&~~=-\frac{\alpha}{\pi}\int_{m_\mu^2}^{Q^2} \frac{d\ell_\perp^2}{\ell_\perp^2}  \ln \frac{Q^2}{\ell_\perp^2}+
\frac{\alpha}{\pi}\int_{m_\mu^2}^{\mu_r^2} \frac{d\ell_\perp^2}{\ell_\perp^2}  \ln \frac{Q^2}{\ell_\perp^2}\nonumber\\
&&~~=- \frac{\alpha}{2\pi}\ln^2\frac{Q^2}{\mu_r^2}, 
\end{eqnarray}
where $\mu_r=c_0/r_\perp$ with $c_0=2e^{-\gamma_E}$ and $\gamma_E$ the Euler constant. In the above calculation, we notice the fact that the real and virtual completely cancel in the region $\ell_\perp <m_\mu$. 

However, when $\mu_r <m_\mu$, we find the virtual contribution reads
\begin{eqnarray}
S_u^{(v)}|_{\mu_r<m_\mu}&=&-\frac{\alpha}{\pi}\int_{m_\mu^2}^{Q^2} \frac{d\ell_\perp^2}{\ell_\perp^2}  \ln \frac{Q^2}{\ell_\perp^2}\nonumber\\
&&-\frac{\alpha}{\pi}\int_{\mu_0^2}^{m_\mu^2} \frac{d\ell_\perp^2}{\ell_\perp^2}  \ln \frac{Q^2}{m_\mu^2} \ ,
\end{eqnarray}
where a lower cutoff $\mu_0$ has been introduced to regulate the infrared divergence. Similarly, for this case, the real contribution also depends on $\mu_0$,
\begin{equation}
S_u^{(r)}|_{\mu_r<m_\mu}=\frac{\alpha}{\pi}\int_{\mu^2}^{\mu_b^2} \frac{dk_\perp^2}{k_\perp^2}  \ln \frac{Q^2}{m_\mu^2} \ .
\end{equation} 
Adding them together, we find that 
\begin{equation}
S_u^{(r)}+S_u^{(v)}|_{\mu_r<m_\mu}=- \frac{\alpha}{2\pi}  \ln \frac{Q^2}{m_\mu^2} \left[\ln \frac{Q^2}{\mu_b^2} +\ln \frac{M^2}{\mu_r^2}\right]\ .
\end{equation}
To summarize the above derivations, we can write the complete one-loop results in the Fourier transform $r_\perp$-space, 
\begin{eqnarray}
&&S_{u}(Q,m_\mu;r_\perp)\\
&&~~=
\begin{cases}
-\frac{\alpha}{2\pi} \ln^2\frac{Q^2r_\perp^2}{c_0^2}\ ,& m_\mu r_\perp< 1 \ , \\
-\frac{\alpha}{2\pi}\ln\frac{Q^2}{m_\mu^2}\left[\ln\frac{Q^2r_\perp^2}{c_0^2}+\ln\frac{m_\mu^2r_\perp^2}{c_0^2}\right]\ ,& m_\mu r_\perp> 1 \ .\nonumber
\end{cases}
\label{su0}
\end{eqnarray}
Since soft photon radiation factorizes, all order resummation can be written as a simple exponential of $S_u(Q,m_\mu;r_\perp)$. 

As discussed above, the derivations can be carried out for the total transverse momentum distribution from soft photon radiation in the Lab frame as well. We will obtain the similar result as that in Eq.~(\ref{soft0}), where $\ell_\perp$ is now replaced by $\vec{p}_T=\vec{p}_{1T}+\vec{p}_{2T}$ in the Lab frame. Again, when Fourier transformed into the $r_\perp$-space, we will have the same result as Eq.~(\ref{su0}), as $r_\perp$ can be regarded as the Fourier conjugate variable for $p_T$ as well.

To compute the final result for the total transverse momentum distribution of the lepton pair, we convolute the above all order Sudakov contribution with the incoming two photons contributions in the coordinate space to ensure momentum conservation. The transverse momentum distribution for the latter can be formulated from the effective photon approach discussed in the previous section. As an example to illustrate the soft photon radiation effects, we assume a simple Gaussian form for the initial two photon contribution, and the total transverse momentum distribution can be written as
\begin{equation}
\frac{dN}{d^2p_T}=\int\frac{d^2r_\perp}{(2\pi)^2}e^{ip_T\cdot r_\perp} e^{-\frac{r_\perp^2 Q_0^2}{4}}e^{-S_u(Q,m_\mu;r_\perp)} \ ,\label{res0}
\end{equation}
where the first term represents the initial two photon contribution as a Gaussian distribution with width $Q_0=40\rm\ MeV$. The second term represents an all order resummation of the soft photon radiation contribution. Although there is no analytic expression for the Fourier transform of the above result in the transverse momentum space, we find the following solution is a very good approximation,
\begin{eqnarray}
&&\int\frac{d^2r_\perp}{(2\pi)^2}e^{ip_T\cdot r_\perp}e^{-S_u(Q,m_\mu;r_\perp)}e^{-\frac{Q_0^2r_\perp^2}{4}} \label{e16}\\
&&~\approx \frac{\Gamma(1-\beta)e^{ \gamma_0}}{\pi Q_0^2}
\left(\frac{Q_0^2}{Q^2}e^{-2\gamma_E}\right)^\beta  {}_1F_1\left(1-\beta,1,-\frac{p_T^2}{Q_0^2}\right)\ ,\nonumber 
\end{eqnarray}
where ${}_1F_1$ is a hypergeometric function, $\gamma_0=\frac{\alpha}{2\pi}\ln^2\frac{Q^2}{m_\mu^2}$, $\beta= \frac{\alpha_e}{\pi}\ln\frac{Q^2}{p_T^2+m_\mu^2}$, and $\gamma_E$ the Euler constant. Numerically, we have checked that the above expression gives very close result for the Fourier transform of Eq.~(\ref{res0}).

\begin{figure}
\includegraphics[width=7cm]{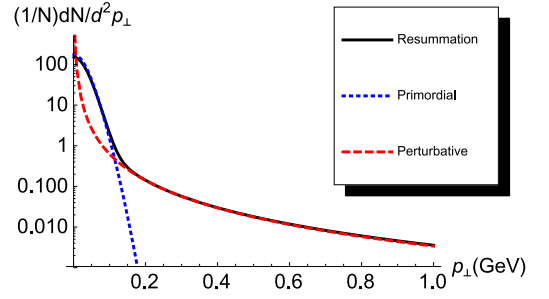}
\caption{Compare the $p_T$ distribution contributions for a typical lepton pair production kinematics: incoming photon flux, perturbative [Eq.(\ref{soft0})], and total contribution with the Sudakov resummation. Leptons are produced at mid-rapidity with invariant mass $10$GeV.}
\label{pt}
\end{figure}

In Fig.~\ref{pt}, we compare the contributions from the incoming photon fluxes (`primordial') and the perturbative photon radiation with the total contribution with Sudakov resummation. First, we notice that the primordial photon flux contribution dies out rapidly around $70\rm\ MeV$, where the perturbative contribution takes over. Second, the Sudakov resummation result is consistent with the one soft photon contribution at relative large transverse momentum. This is understandable, because the electromagnetic coupling $\alpha_e$ is small, and the resummation result is dominated by the first order corrections in this region. On the other hand, between these two regions, the resummation result provides a smooth match between the primordial distribution and one photon radiation contribution. 

It is interesting to compare the distributions of Fig.~\ref{pt} to the Drell-Yan type lepton pair production, where the lepton pair are produced through quark-antiquark annihilation process. In the Drell-Yan process, not only the incoming quark distributions but also the QCD Sudakov effects contribute to a significantly higher transverse momentum. The latter can be seen from the invariant mass dependence of the $p_T$ spectrum. In particular, for high mass final states, like $Z$-boson production, the low $p_T$ spectrum are overwhelmingly dominated by the Sudakov effects, see, e.g., the discussions in Ref.~\cite{Su:2014wpa}. Numerically, for the same invariant mass range of lepton pair, the $p_T$ distribution is peaked around few GeV for Drell-Yan process~\cite{Su:2014wpa}, which is order of magnitude higher than the spectrum of the pure QED process shown in Fig.~\ref{pt}. 

\subsection{Comparison with the UPC Data from ATLAS}

Combining the Sudakov resummation of all order soft photon radiation with the incoming photon fluxes contribution, we have the following expression for the total transverse momentum of the lepton pair in heavy ion collisions,
\begin{eqnarray}
\frac{d\sigma(AB_{[\gamma\gamma]}\to\mu^+\mu^-)}{dy_1dy_2d^2p_{1T}d^2p_{2T}}&\!=\!&\!\sigma_0\!\int\!\frac{d^2r_\perp}{(2\pi)^2}e^{ip_T\cdot r_\perp}W(b_\perp;r_\perp) \ ,
\end{eqnarray}
where $\sigma_0$ is defined in Eq.~(\ref{e4}). $W(b_\perp;r_\perp)$ contains the incoming photon flux and all order Sudakov resummation,
\begin{eqnarray}
W(b_\perp;r_\perp)={\cal N}_{\gamma\gamma}(b_\perp;r_\perp)e^{-S_u(Q,m_\mu;r_\perp)}
\ ,\label{res}
\end{eqnarray}
where $S_u$ is defined in Eq.~(\ref{su0}). We have introduced a short notation ${\cal N}$ for the incoming photon flux contribution. 

As mentioned in the introduction, in the experiments, the azimuthal angular correlation has been commonly applied to study the low transverse momentum behavior of two final state particles. When the total transverse momentum is relatively small, the two leptons in the final state are almost back-to-back in the transverse plane with small angle $\phi_\perp=\pi-\phi_{12}$. We are interested in the small $\phi_\perp\ll \pi$ region. For convenience, we take one of the leptons' transverse momentum as reference for $-\hat x$ direction, $p_{1T}\sim (-|P_T|,0)$, and $p_{2T}$ is parameterized by $p_{2T}=(|P_T|\cos\phi_\perp,|P_T|\sin\phi_\perp)$. The total transverse momentum is $p_T$. In the ATLAS experiment, the measurements are presented as functions of the so-called acoplanarity $\alpha$, which is defined as $\alpha=|\phi_\perp|/\pi$. 

In order to compare to the UPC events from ATLAS experiment~\cite{ATLAS:2016vdy}, we apply the following assumptions to simplify the numeric calculations. First, we apply the average photon flux for each nucleus to represent the relative photon flux as function of $x_a$ and $x_b$, respectively. Much of the uncertainties introduced by this assumption will be cancelled out in the normalized distributions when we compare to the ATLAS data. In particular, we have compared the invariant mass and rapidity distributions of the lepton pair to the ATLAS measurements for the UPC events, and found very good agreements. Second, we assume a simple Gaussian distribution for the transverse momentum dependence in the incoming photon flux. The Gaussian width for each photon flux is computed from Eq.~(\ref{nucleus}) as function of $x_{a,b}$. With the above approximations, we have the following expression for ${\cal N}$,
\begin{eqnarray}
{\cal N}_{\gamma\gamma}(b_\perp;r_\perp)&\approx &\left[x_af_A^\gamma(x_a)x_bf_B^\gamma(x_b)\right]\nonumber\\
&&\times e^{-\frac{\left(Q_0^2(x_a)+Q_0^2(x_b)\right)r_\perp^2}{4}} \ .
\end{eqnarray}
In the above equation, $Q_0(x_a)$ and $Q_0(x_b)$ represent the average transverse momentum for the photon fluxes of two incoming nuclei, respectively. For the typical kinematics of ATLAS measurements~\cite{ATLAS:2016vdy}, we find that the average $Q_0^2(x_a)+Q_0^2(x_b)=\left(40{\rm MeV}\right)^2$ for $x_a=x_b=10^{-3}$. This is consistent with the results shown in Fig.~\ref{avektAB} as well.

\begin{figure}
\includegraphics[width=7cm]{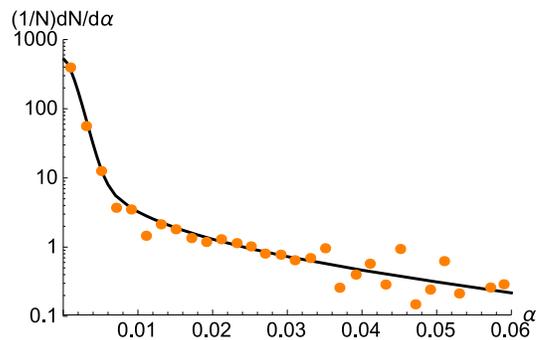}
\caption{Acoplanarity distribution for lepton pair production in UPC events at the LHC~\cite{ATLAS:2016vdy} for invariant mass from 10 to 100 GeV and rapidity from -0.8 to 0.8. There is a $10\%$ normalization uncertainty in the comparison, and the unfolding effects from the experiments are not taken into account either.}
\label{upcalpha}
\end{figure}

In Fig.~\ref{upcalpha}, we show this comparison for the typical kinematics of the ATLAS measurement of UPC events: lepton transverse momentum $P_T>4\ \rm GeV$, rapidity $|Y_{\mu\mu}|<0.8$, and invariant mass $10{\rm GeV}<M_{\mu\mu}<100{\rm\ GeV}$. This figure shows that the theoretical results agree with the ATLAS measurements very well. This provides an important baseline for the central collisions, for which we will discuss the following section.

There have been suggestions that the so-called the nucleus dissociation contribution may be important at relative large transverse momentum, where the individual proton in one of nucleus contributes incoherently to the production process. However, in this region, the incoherent photon flux depends on the proton form factor, whose power behavior at large transverse momentum leads to a much smaller contribution as compared to the soft photon and resummation contribution. 

\section{Medium Interactions with Leptons}

In previous sections, we focused on UPCs. In a recent experiment measurement, the ATLAS collaboration has extended this idea to central collisions. It was argued that the two photon scattering processes produce di-lepton with small total transverse momentum (which is similar to UPC events), and these processes probe the electromagnetic property of the quark-gluon plasma when they traverse through the medium in the central collision events. In particular, the lepton pair has very small transverse momentum from two-photon scattering processes (see the discussions in the last section), the medium effects can be evidently measured through the so-called acoplanarity. 

Before we deal with medium effects in lepton pair production in central $AA$ collisions, we would like to comment on the contribution from partonic photon-photon scattering.  This comes from the photon distribution functions from the nucleons in both nuclei. The differential cross section can be written as those in previous sections, and the Sudakov resummation can be derived as well. The only difference is that we have to involve the parton distributions for the photons from the nucleons~\cite{Manohar:2016nzj,{Manohar:2017eqh}}. Essentially, the photon distribution function is calculated from the quark distribution. Therefore, the photon PDF contribution scales with $A$, whereas that the photon flux contribution scales with $A^2$ as discussed in Sec.~II. In addition, for the relevant kinematics of our study, i.e., low transverse momentum region, we have to apply the nucleon form factor for the photon PDFs, which is of order $\Lambda_{QCD}$. It is much larger than the typical momentum region of this study, see, plots in Fig.~\ref{pt}. Therefore, we can safely ignore the photon PDF contributions. 

\subsection{Inclusion of Medium Effects}

The approach is very similar to that of the dijet azimuthal angular correlation in heavy ion collisions, where the $P_T$ broadening of energetic jets in hot QCD medium can be investigated. However, in the LHC energy range, the dominant broadening effects actually comes from the Sudakov effect~\cite{Mueller:2016gko} for the typical dijet kinematics there~\cite{Aad:2010bu,Chatrchyan:2011sx}. From the analysis in the previous section, we find that for the lepton pair production through two-photon scattering process, it is completely opposite. The QED Sudakov effect is dominated by the incoming photon transverse momentum in a wide range of kinematics. If the medium effects is strong enough, we will be able to probe it by measuring the azimuthal angular correlations between the lepton pair.

\begin{figure}
\includegraphics[width=7cm]{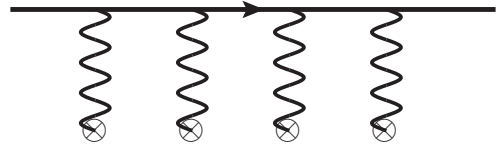}
\caption{Multiple interactions of the energetic lepton with the medium, where the symbol $\otimes$ represents the scattering center from the charged medium.}
\label{qedpt}
\end{figure}

Because the lepton only carries electric charge (we ignore the weak charge), it only interacts with the electrically charged particles in the medium. As illustrated in Fig.~\ref{qedpt}, the lepton suffers multiple interactions with the medium during this process. We can resum the multiple photon exchanges between the produced lepton and the electric charged particles in the medium, and find that this results in a QED type time-ordered Wilson line as follows
\begin{eqnarray}
\mathcal{U}_{\textrm{QED}} (x_\perp) &=&T\exp \bigg[-ie\int dz^- \int d^2 z_\perp G(x_\perp-z_\perp) 
\nonumber\\
&&~\times 
\rho_e (z^-, z_\perp)\bigg], 
\end{eqnarray}
where $\rho_e (z^-, z_\perp)$ is the random electric charged density of the target medium, and the photon propagator $G(x_\perp)$ is defined as 
\begin{equation}
G(x_\perp)= \int \frac{d^2q_\perp}{(2\pi)^2}  \frac{1}{q_\perp^2+\lambda^2}e^{iq_\perp\cdot x_\perp}=\frac{1}{2\pi}\mathrm{K}_0(\lambda x_\perp)\ ,
\end{equation}
with $\lambda$ acting as an infrared (IR) regulator such as the Debye mass in QED medium. The value of $\lambda$ can be estimated according to the range of the electromagnetic interaction in the medium, which depends on the QED Debye screening mass $m_D \sim \frac{eT}{\sqrt{3}}$, where $T$ is the temperature. In our case, simple estimates show that $m_D \sim 80 \, \text{MeV}$ which corresponds to a Debye length a few times less than the typical size of the medium created in central heavy ion collisions. 

Similar to the QCD $q\bar q$ dipole calculation, the QED incoherent multiple scattering amplitude between the $\ell^+\ell^-$($q\bar q$) dipole with size $r_\perp$ and the target medium can be cast into
\begin{equation}
\langle\mathcal{U}_{\textrm{QED}} (b_\perp+\frac{1}{2}r_\perp) \mathcal{U}^{\dagger}_{\textrm{QED}} (b_\perp-\frac{1}{2}r_\perp)\rangle =\exp\left[-\frac{Q_{se}^2r_\perp^2}{4}\right], 
\end{equation}
where the analog of saturation momentum in QED $Q_{se}^2\equiv \frac{e^4}{4\pi}\ln \frac{1}{\lambda^2 r_\perp^2}\int dz^- \mu_e^2 (z^-)$. Here, $\mu_e^2$ is related to the expectation of the local charge density fluctuations. As we can see, only a logarithmic dependence on $\lambda$ is left in the above dipole amplitude due to cancellation. Normally one does not have to take into account the multiple scattering for the QED calculation. On the other hand, the dipole size $r_\perp \sim 1/q_\perp$ is sufficiently large in the soft momentum transfer region, which makes $Q_{se}^2 r_\perp^2 \sim 1$. Similar to the QCD case, the correlation of two charge density follows 
\begin{equation}
\langle \rho_e (z^-,z_\perp)\rho_e (z^{\prime -},z_\perp^\prime) \rangle =\delta (z^{-}-z^{\prime -})\delta^{(2)}(z_\perp -z_\perp^\prime) \mu_e^2(z^-).
\end{equation}
In comparison, we often define the QCD saturation momentum as follows
\begin{equation}
Q_{sg}^2 =\frac{N_c}{C_F} Q_{sq}^2\equiv \frac{N_c g^4}{4\pi}\ln \frac{1}{\Lambda^2 r_\perp^2}\int dz^- \mu_c^2 (z^-),
\end{equation}
where $\int dz^- \mu_c^2 (z^-)
= \frac{A}{2\pi R^2}$ is related to the color charge density with $R$ being the size of the target medium\cite{Iancu:2002xk}, and $\Lambda$ is the QCD Debye mass. The above expression is equivalent to the saturation momentum expression for cold nuclear matter $Q_s^2 =\frac{8\pi^2 \alpha_s N_c}{N_c^2-1}\rho\sqrt{R^2-b^2}xG(x,1/r_\perp^2)$\cite{Mueller:1999wm}, where $xG(x,1/r_\perp^2)=N_c\frac{\alpha_s C_F}{\pi}\ln \frac{1}{\Lambda^2 r_\perp^2}$.

\subsection{Implication from the ATLAS Measurements}

If we assume the multiple scattering between the produced di-lepton and the medium similar to the QCD case, we can modify the above $W(r_\perp)$ of Eq.~(\ref{res}) as,
\begin{eqnarray}
W(b_\perp;r_\perp)={\cal N}_{\gamma\gamma}(b_\perp;r_\perp)e^{-S_u(Q,m_\mu;r_\perp)}
e^{-\frac{\langle\hat q_{QED}L\rangle r_\perp^2}{4}}\ ,\nonumber\\
\end{eqnarray}
where the last factor comes from the medium contribution to the di-lepton $p_T$-broadening effects and $\hat q_{QED}$ represents the electromagnetic transport coefficient of the quark-gluon plasma. $\langle \hat q_{QED} L\rangle$ can be identified as the saturation momentum $Q_{se}^2$ discussed above.

\begin{figure}
\includegraphics[width=7cm]{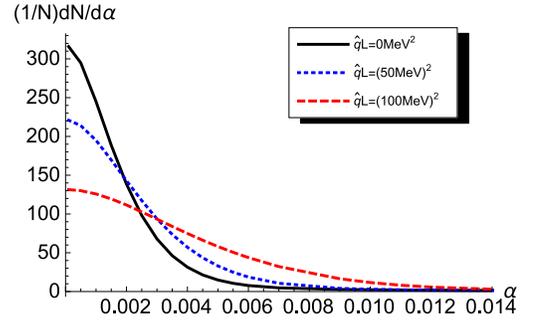}
\caption{Medium modifications to the Acoplanarity distribution, with different values of the effective $\hat qL$.}
\label{central}
\end{figure}

With this additional $p_T$-broadening effects, we can calculate the azimuthal angular correlation in heavy ion collisions. We assume that the initial distributions remain the same as the UPC case and take in account the additional broadening effects by parameterizing the $\hat q L$ for the leptons. 

In Fig.~\ref{central}, using the UPC as the baseline and assuming the mild $b_perp$ dependence due to the initial transverse momentum broadening, we show this effects by imposing several different values of the $\hat qL$. Comparing these curves to the ATLAS measurements, we find that the effective $\langle \hat q_{QED}L\rangle$ range from $(100{\rm MeV})^2$ in most central collisions to $(50{\rm MeV})^2$ in non-central collisions. In the above-mentioned GEPA model, we expect that the effective value of $\langle \hat q_{QED}L\rangle$ becomes smaller. The determination of $\langle \hat q_{QED}L\rangle$ requires a more sophisticated and detailed comparison with the accurate unfolded data.

In order to extract the $p_T$-broadening parameter for the leptons, we need to implement the realistic geometry of the collisions and the photon sources. From that, we shall estimate the average length of the leptons traversing through the medium. Compared to the dijet correlation in heavy ion collisions, this is even more important because the photon source profile is different from the medium profile created in heavy ion collisions. We leave the detailed studies in a future publication. 

\subsection{Comparison with the QCD $p_T$-broadening of Jets: Parametric Estimate}

Assuming that the multiple interactions with the medium can be applied to describe both QED for the leptons and the QCD for the quarks and gluons, we can compare the sizes of the $p_T$-broadening effects from the same formalism, i.e., the BDMPS formalism~\cite{Baier:1996vi,{Baier:1996sk}, Peigne:2008wu}.

Because the leptons have large transverse momenta, their multiple interactions with the medium can be formulated following the BDMPS~\cite{Baier:1996vi,Baier:1996sk} formalism. In particular, the $p_T$-broadening can arise from multiple interaction with the charged particles in the medium. This is quite the same as that for the quark and gluon jets traversing the quark-gluon plasma medium, which also suffer $p_T$-broadening because of multiple QCD interactions with the medium. Of course, the major difference is that the lepton interacts with the electric charges while the quark/gluon jets interact with the color charges. Here, we make some parametric estimates based on this idea.

First, the coupling constant plays an important role. By definition, the $p_T$-broadening depends on two orders of coupling constants. Since the couplings in QED and QCD are dramatically different, thus there is a major difference for the $p_T$-broadening effects. 

Second, the broadening depends on the charge density of the medium. Therefore, we need to estimate the density difference for the electromagnetic QED and strong interaction QCD densities in the quark-gluon plasma. Since only quarks carry electric charges, we can determine the electric charge density from the quark density. For the strong interaction case, both quarks and gluons contribute. The parton density is proportional to the degree of freedom if we assume the thermal distributions of the quarks and gluons. Therefore, the ratio of quark density vs the gluon density goes as, see, for example, Ref.~\cite{Blaizot:2014jna}: $\frac{21}{2}N_f: 16$, where $N_f$ is the number of active flavors. There are two additional differences: one is the color factor in QCD case, and the other is the charge average between the quarks and gluons. For electric charge average, we have
\begin{equation}
    \sum_{u,d,s} e_q^2= \frac{2}{9} N_f\ , 
\end{equation}
where we take $N_f=3$ with active $u$, $d$ and $s$ quarks. For the color factor associated with the quark density, it can be evaluated as 
\begin{equation}
    \frac{1}{N_c^2}{\rm Tr}[T^aT^b]{\rm Tr}[T^aT^b]=\frac{2}{9} \ ,
\end{equation}
which is an effective color factor by considering the quark-quark scattering amplitude, where $N_c=3$. For gluon density case, it is
\begin{equation}
    \frac{1}{N_c(N_c^2-1)}f_{abc}f_{abd}{\rm Tr}[T^cT^d]=\frac{1}{2}
\end{equation}
from the quark-gluon scattering amplitude. These are for the quark jet $p_T$-broadening.  Therefore, we can estimate the ratio between the QED and QCD saturation scales as 
\begin{equation}
\frac{\langle \hat q_{QED}L\rangle }{\langle\hat q_{QCD}L\rangle}=\frac{\alpha_e^2}{\alpha_s^2}\frac{\frac{21}{2}N_f\frac{2}{9}}{\frac{21}{2}N_f\frac{2}{9}+16\frac{1}{2}}= \frac{\alpha_e^2}{\alpha_s^2}\times\frac{7}{15}\ . \label{ratio}
\end{equation}
where again $N_f=3$. Here $\langle \hat q L\rangle$ represents the amount of medium broadening, which is known as the saturation scale in the so-called dipole formalism. The above estimate of QCD $\langle \hat q L\rangle$ is for QCD quark jets. For gluon jets, a factor of $C_A/C_F$ should be multiplied to the denominator. 

We would like to add a few comments about these observations. One is about the thermal density ratio we applied above. We assume that quark and gluons are thermalized at the same time, which may not be true. There has been concern that the quarks may be thermalized at a later time, see, for example, the discussions in Ref.~\cite{Blaizot:2014jna}. Second, we did not take into account the details of the medium property, for example, the associated Debye masses for QED and QCD. This could introduce additional complexity in the calculations. In addition, for the QCD case, there are length dependent double logarithms~\cite{Liou:2013qya}. Last but not least, the medium path length $L$ can differ for the QED and QCD cases, since the electron pair can be created outside the medium. If all these effects are taken into account, Eq.~(\ref{ratio}) may not apply. Nevertheless, the above equation can serve as a simple formula for a rough estimate. 

\subsection{Comments on the QED energy loss}

In QED medium, the radiative energy loss can be estimated as~\cite{Baier:1996vi},
\begin{equation}
\Delta E_{QED}=\sqrt{\langle \hat qL\rangle}\frac{\alpha}{\pi}\frac{2}{3}\sqrt{\frac{L^2E}{\lambda_m}}
\end{equation}
in a certain kinematic limit of the induced radiated photon spectrum, where $L$ represents the average path of the lepton in the medium, $\lambda_m$ for the mean field path of the medium. For the kinematics of our interest, we find that the total energy loss is a few percent of the $p_T$-broadening size ($100\rm MeV$). It is so small that we will not be able to observe such effects.

The QED case is totally different from QCD jet quenching in heavy ion collisions. First, the energy loss of the leptons is much smaller than the QCD jet energy loss. For the leptons going through the medium, there are no surface bias effects, because the energy loss is completely negligible. However, we know that the QCD jet energy loss is important, and surface bias effects are significant in the realistic simulation of jet physics in heavy ion collisions. 

\section{Magnetic Field Effects}

One measurement  might indicate that the $p_T$-broadening effects of the di-lepton could come from the magnetic effects created at very early time of heavy ion collisions~\cite{Adam:2018tdm}. The major phenomenological difference between this mechanism and the multiple scattering mechanism discussed in this paper is that the magnetic effects is strongly correlated with the event plane. Therefore, if we can measure this correlation, we could distinguish these two mechanisms.

In addition, the contribution from the magnetic effects is a cross product: $\vec{B}\times \vec{V}$, where $\vec{B}$ is the magnetic field generated in heavy ion collisions and $\vec{V}$ is the lepton velocity. This introduces two important observational consequences: (1) impact parameter dependence of the effects is different from the multiple interaction with the medium discussed in this paper, which will increase with decreasing impact parameter. For the magnetic effects, it increases from UPC to peripheral collisions but will decrease for more central collisions with decreasing impact parameter. (2) The magnetic effects for the additional $p_T$-broadening happens in the direction perpendicular to both the magnetic field $\vec{B}$ and the lepton moving direction. Therefore, if both the lepton and anti-lepton have the same rapidity and they have almost same size of transverse momentum as in our case, the magnetic effects will cancel out between them. That is, the lepton and anti-lepton will gain same amount of additional transverse momentum kick from the magnetic effects but with opposite signs. On the other hand, if the lepton and anti-lepton are back-to-back in the $\hat z$-direction, i.e., their rapidities are opposite to each other, the magnetic fields will have net effects on the transverse momentum kick and lead to the observed transverse momentum broadening in the lepton pair. 

Therefore, we should be able to distinguish these two mechanisms by studying the centrality dependence and rapidity dependence of the $p_T$-broadening effects. In the following, we will discuss the details of the magnetic effects. First, we show that the contributions from initial electromagnetic fields generated by the colliding nuclei cancel out completely in the leading power of the lepton pair production through the two-photon process. Then, we will discuss how to measure the residual magnetic field (due to the created quark-gluon plasma) effects.

\subsection{Initial Electromagnetic Fields Contributions: the classical electrodynamics perspective}

To estimate the electromagnetic fields of a fast moving nucleus, let us first follow the approach in classical electrodynamics~\cite{Jackson:1998nia}. Then we provide some discussion and interpretation in terms of quantum field theory calculations. We can write the electromagnetic fields of a moving charge propagating along the $z$ direction in the lab frame as 
\begin{equation}
E_x= \frac{\gamma q b}{(b^2+\gamma^2 v^2 t^2)^{3/2}}, \quad B_y=\beta E_x,
\end{equation}
where $\gamma =\frac{1}{\sqrt{1-\beta^2}}$, $\beta=v/c$ and we put the impact parameter $b$ in the $x$ direction. Let us compute the electromagnetic field effect for one particular nucleus and choose such a frame that the velocities of the back-to-back $\mu^+$ and $\mu^-$ can be written as $v_+ =(v_x, v_y, v_z)$ and $v_- =(-v_x, -v_y, -v_z)$, respectively. In the Eikonal approximation, we can find the force in the $y$-direction is zero, and the Lorentz forces in the $x$-direction for $\mu^+$ and $\mu^-$ can be written as   \begin{eqnarray}
\frac{dp_{+x}}{dt}&=&eE_x (1-\beta_z \beta), \\
\frac{dp_{-x}}{dt}&=& -eE_x (1+\beta_z \beta),
\end{eqnarray}
respectively. Also, we need to include the effect of time dilation since $\mu^+$ and $\mu^-$ are moving along the $z$ direction with finite velocities. 
The relative velocities are
\begin{equation}
\beta_+ = \frac{\beta_z - \beta}{1 -\beta_z  \beta}, \quad  \beta_- =- \frac{\beta_z + \beta}{1 +\beta_z  \beta}.
\end{equation}
Therefore, we find the times that $\mu^+$ and $\mu^-$ experience the electromagnetic shockwave are different and their ratio is 
\begin{equation}
\frac{\Delta t_+}{\Delta t_-} =\sqrt{\frac{1-\beta_+^2}{1-\beta_-^2}}.
\end{equation}
It is then straightforward to check that 
\begin{eqnarray}
\Delta p_{+x} &=& eE_x \Delta t (1-\beta_z \beta) \sqrt{1-\beta_+^2} \nonumber\\
&=&eE_x \Delta t \sqrt{1+\beta^2\beta_z^2-\beta^2-\beta_z^2} \nonumber\\
&=& -\Delta p_{-x} .
\end{eqnarray}
This demonstrates that the momentum broadening exactly cancel for exact back-to-back lepton pairs due to charge neutrality. To get a non-zero transverse momentum broadening, we need to take the charge dipole moment into account. In coordinate space, the separation of the lepton pair is proportional to $1/Q$, where $Q$ is the invariant mass of the produced lepton pair. Therefore, the total transverse momentum broadening square must be proportional to $\langle P_m^2 \rangle\frac{1}{Q^2R_A^2}$, where $\langle P_m^2 \rangle \sim e^2E_x^2 R_A^2\sim \frac{Z^2e^4}{R_A^2} \sim (30\textrm{MeV})^2$, where $R_A\approx 7\ \rm fm$ is the radius of the nucleus. 

Here we have used a quantum-classical mixed picture to illustrate the broadening of the 
produced lepton pair due to the coherent electromagnetic fields of fast moving nuclei by 
assuming that the lepton pair is created before the pair passes over the high energy nuclei. 
This intuitive picture helps to estimate and understand the final state broadening of the lepton pair,
while a fully quantum treatment of this process is presented in the next subsection. 

\subsection{Initial Electromagnetic Fields Contributions: the quantum field theory perspective}

Quantitatively, we can also study the $p_T$-broadening effect in terms of Wilson lines from the quantum field theory perspective. Let us derive the QED Wilson by using the so-called Lienard-Wiechert potential for a moving point charge particle with charge $q=Ze$. Again, we can first follow the discussion and the convention in Ref.~\cite{Jackson:1998nia}, which gives the following four vector potential 
\begin{equation}
A^{\alpha}(x) =\frac{qV^{\alpha}(\tau)}{V\cdot [x-r(\tau)]}|_{\tau=\tau_0}\ .
\end{equation}
Here $V^{\alpha}=(\gamma c, \gamma c\beta)$ is the four vector velocity of the moving charge. Putting in the light-cone constraint $\tau=\tau_0$, and using Eq.(14.16) together with the fact $d x_\alpha =V_\alpha d\tau =V_\alpha \gamma dt$, 
we can write
\begin{equation}
\int  A^{\alpha} (x) dx_{\alpha} =q \int_{-\infty}^{\infty}  \frac{\gamma dt}{\sqrt{b_\perp^2+\gamma^2 v^2 t^2}}  \ .
\end{equation}
The above integration is divergent, but it becomes finite if we consider the difference of two opposite charges which are separated by a distance $r_\perp$. Therefore, 
the result for a lepton pair lepton is
\begin{eqnarray}
&&e\int  A^{\alpha} (x) dx_{\alpha}- e\int  A^{\alpha} (x+r_\perp) dx_{\alpha} =eq \int_{-\infty}^{\infty} \gamma dt\nonumber\\
&&~\times\left[  \frac{1}{\sqrt{b_\perp^2+\gamma^2 v^2 t^2}} - \frac{1}{\sqrt{(b_\perp+r_\perp)^2+\gamma^2 v^2 t^2}}\right].
\end{eqnarray}
Eventually, if we set $v=c=1$, we can obtain 
\begin{eqnarray}
&&e\int  A^{\alpha} (x) dx_{\alpha}- e\int  A^{\alpha} (x+r_\perp) dx_{\alpha}  \nonumber\\
&&~~~~~=eq \ln \frac{(b_\perp+r_\perp)^2}{b_\perp^2}\ .
\end{eqnarray}
By taking into account the fact that $q=Ze$ for a relativistic heavy ion, and the $\frac{1}{4\pi}$ difference due to different conventions, we can see that the above result agrees completely with the Wilson line formalism used below. 

Next, in terms of the final state QED type scattering and Wilson lines approaches \cite{Ivanov:1998ka, Brodsky:2002ue, Xiao:2010sa}, we can compute the cross section $\sigma_{\gamma A\to \mu^+\mu^- A} $ as follows. It is straightforward to resum the multiple photon exchanges between a muon line and target nuclei, and find that this results in a QED type Wilson line 
\begin{equation}
\mathcal{U}_{\textrm{QED}} (x_\perp) =\exp \left[iZe^2 G(x_\perp)\right], 
\end{equation}
with the photon propagator $G(x_\perp)$ defined as 
\begin{eqnarray}
G(x_\perp)
=\frac{1}{2\pi}\mathrm{K}_0(\lambda x_\perp).
\end{eqnarray}
Here we use $\lambda$ as the IR cut off for the moment. When we study the muon pair production, the $\lambda$ dependence will cancel. In this simple model, we assume that the nuclei is a point particle with charge $Ze$ in order to study the coherent final state effect. The QED multiple scatterings contribute a phase to the QED type $\mu^+\mu^-$ dipole with size $r_\perp$ as follows
\begin{eqnarray}
&&\mathcal{U}_{\textrm{QED}} (b_\perp+\frac{1}{2}r_\perp) \mathcal{U}^{\dagger}_{\textrm{QED}} (b_\perp-\frac{1}{2}r_\perp)\nonumber\\
&&~~~=\exp\left[2iZ\alpha \ln\frac{|b_\perp+\frac{1}{2}r_\perp|}{|b_\perp-\frac{1}{2}r_\perp|}\right].
\end{eqnarray}
Furthermore, let us use $\sigma_{\gamma A\to \mu^+\mu^- A} $ as an example to estimate the order of magnitude of the electromagnetic corrections in the final state between the produced muon pair and the target nucleus with charge number $Z$. The cross section can be written as 
\begin{eqnarray}
&&\frac{d\sigma_{\gamma A\to \mu^+\mu^- A}}{d^2 p_T d^2P_T} = N \int \frac{d^2 b_\perp}{(2\pi)^2} \frac{d^2 b_\perp^\prime }{(2\pi)^2} \frac{d^2 r_\perp }{(2\pi)^2} \frac{ d^2 r_\perp^\prime }{(2\pi)^2} \nonumber\\
&&~~\times \psi (r_\perp) \psi ^\ast (r_\perp^\prime) e^{-ip_T\cdot (b_\perp -b_\perp^\prime ) -iP_T\cdot (r_\perp -r_\perp^\prime ) }\nonumber\\
&&~~\times \left[1-\exp\left(2iZ\alpha \ln\frac{|b_\perp+\frac{1}{2}r_\perp|}{|b_\perp-\frac{1}{2}r_\perp|}\right) \right]   \notag \\
&&~~ \times \left[1-\exp\left(-2iZ\alpha \ln\frac{|b^\prime_\perp+\frac{1}{2}r_\perp^\prime|}{|b_\perp^\prime-\frac{1}{2}r_\perp^\prime|}\right) \right],
\end{eqnarray}
where the splitting function $\psi (r_\perp) \sim \frac{\epsilon\cdot r_\perp}{r_\perp^2}$ represents the contribution from the $\gamma \to \mu^+\mu^-$ splitting. Also if we expand the expression $\left[1-\exp\left(2iZ\alpha \ln\frac{|b_\perp+\frac{1}{2}r_\perp|}{|b_\perp-\frac{1}{2}r_\perp|}\right) \right] $ to the lowest order, we should be able to recover the lowest order $\gamma \gamma \to \mu^+\mu^-$ cross section together with the proper normalization $N$. In the back-to-back limit, it follows that $P_T^2 \gg p_T^2$. In the case of interest, $P_T$ is around several GeV while $p_T$ is of the order of $\Lambda_m=\frac{1}{R_A}\sim 30$ MeV. 

Let us comment on the above result. First of all, we can compute the Born contribution by expanding the above Wilson line contribution to the lowest order. In the back-to-back limit, we can approximately write
\begin{equation}
1-\exp\left(2iZ\alpha \ln\frac{|b_\perp+\frac{1}{2}r_\perp|}{|b_\perp-\frac{1}{2}r_\perp|}\right) \simeq -i2 Z\alpha \frac{b_\perp\cdot r_\perp}{b_\perp^2}.
\end{equation}
It is then straightforward to estimate that the Born contribution is of the order of $N  \frac{Z^2 \alpha^2}{p_T^2 P_T^4}$.

Second, we can estimate the contribution with one more photon exchange in the amplitude level, which means the expansion of the Wilson line to the second order. Using the fact that the lower bound of the $b_\perp$ integral is roughly $R_A$, we can find that the first final state interaction correction is of the order 
\begin{equation}
 N  \frac{Z^2 \alpha^2}{p_T^2 P_T^4} \frac{Z^2 \alpha^2}{P_T^2 R_A^2} \sim N  \frac{Z^2 \alpha^2}{p_T^2 P_T^4} \frac{Z^2 \alpha^2 \Lambda_m^2}{P_T^2 }. 
\end{equation}
Therefore, the final state interaction is power suppressed by the factor of $\frac{Z^2 \alpha^2 \Lambda_m^2}{P_T^2 }$. Also, since the transverse momentum square average $\langle p_T^2\rangle \sim \Lambda_m^2 \sim \frac{1}{R^2_A}$ at lowest order, we can see that the change due to final state interaction is roughly $\Delta \langle p_T^2\rangle \sim \Lambda_m^2 \frac{Z^2 \alpha^2 \Lambda_m^2}{P_T^2 }$, which is in agreement with our estimate in previous subsection from the perspective of classical electrodynamics. 

In summary, the contributions from the initial electromagnetic fields generated by the colliding nuclei cancel out completely in the leading power of $p_T/P_T$, while the finite contribution to the transverse momentum broadening is power suppressed. This cancellation is also consistent with a factorization argument that the final state interaction effects vanishes in this process due to the opposite charges of the lepton pair.

\subsection{Residual Magnetic Field Effects in the Quark-gluon Plasma}

In the meantime, as argued in some recent papers~\cite{Kharzeev:2009pj,Asakawa:2010bu,Skokov:2016yrj}, there could be a residual and strong magnetic field in the quark-gluon plasma after the heavy ion collisions. Due to the collision symmetry, the magnetic field only contains the perpendicular component $\vec{B}_\perp$. As mentioned at the beginning of this section, since the Lorentz force vanishes along the direction of the magnetic field, the amount of the $p_T$-broadening from the magnetic effects will have a non-trivial correlation with the event plane, which is correlated with the direction of the magnetic field~\cite{Kharzeev:2009pj,Asakawa:2010bu,Skokov:2016yrj}. 

\begin{figure}
\includegraphics[width=6cm]{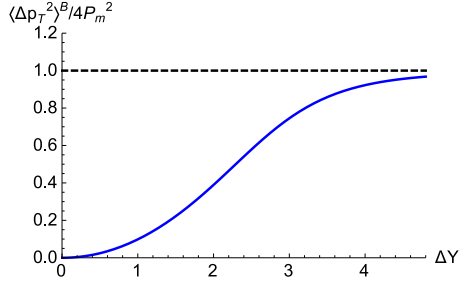}
\caption{Normalized magnetic effects on the $p_T$-broadening for the lepton pair as function of their rapidity difference $\Delta Y=|y_{\mu^+}-y_{\mu^-}|$ with $|y_\mu|<2.4$. }
\label{magneticdy}
\end{figure}

A number of further observations regarding the effects of the residual magnetic fields are as follows. First, the transverse momentum kick comes from the longitudinal motion $v_z$ of the leptons. Therefore, the magnetic effects crucially depend on the rapidity of the leptons. In particular, if the lepton and the anti-lepton move in the same direction, the magnetic effects will cancel out in the total transverse momentum for the pair. Only if they are moving in opposite $\hat z$ direction, there will be net effects. In more detail, the net transverse momentum kick for the pair is proportional to 
\begin{equation}
|\Delta p_y^m|_{\mu^+\mu^-}={\rm \bf P_m} \left[\tanh (y_+)-\tanh(y_-)\right] \ ,
\end{equation}
where $ {\rm \bf P_m}$ represents the average transverse momentum kick depending on the geometry of the collisions, $y_+$ and $y_-$ are rapidities for the lepton and the anti-lepton, respectively. Therefore, the total $p_T$-broadening effects for the pair can be formulated as
\begin{equation}
\langle \Delta p_T^2\rangle_{\mu^+\mu^-}^{B}=\langle {\rm \bf P_m^2}(b_\perp)\rangle \left[\tanh (y_+)-\tanh(y_-)\right]^2 \ ,
\end{equation}
where $ \langle {\rm \bf P_m^2}(b_\perp)\rangle$ stands for the average $p_T$-broadening with a centrality dependence. Assuming that a flat distribution for the rapidity dependence for the two leptons in the range of $|y_\mu|<2$, and the net magnetic effects can be calculated as function of rapidity difference between the lepton pair $\Delta Y=|y_{\mu^+}-y_{\mu^-}|$,
\begin{eqnarray}
&&\langle \Delta p_T^2\rangle_{\mu^+\mu^-}^{B}(\Delta Y)=\langle{\rm \bf P_m^2(b_\perp)}\rangle \\
&&~~\times\frac{\int_{\Delta Y} dy_1 dy_2\left[\tanh (y_1)-\tanh(y_2)\right]^2d\sigma(y_1,y_2)}{\int_{\Delta Y} dy_1 dy_2d\sigma(y_1,y_2)}\ ,\nonumber
\end{eqnarray}
where the rapidity integrals are constrained with $\Delta Y=|y_1-y_2|$. In Fig.~\ref{magneticdy}, we plot the normalized distribution as function $\Delta Y$ for a typical transverse momentum for the lepton $P_T=6\ {\rm GeV}$. From this plot, we can clearly see that the magnetic effects on the $p_T$-broadening increases with rapidity difference, and approach to the maximum values around $\Delta Y=3$. On the other hand, the multiple interaction effects with the medium remain a relative constant because the medium density does not change in this rapidity range. Therefore, the difference between the amount of the $p_T$-broadening effects at different $\Delta Y$ can serve as an effective measure for the magnetic effects:
\begin{equation}
    \left[\langle\Delta p_T^2\rangle_{\Delta Y=3}-
        \langle\Delta p_T^2\rangle_{\Delta Y=0}\right]_{b_\perp}\propto \langle \vec{B}_\perp^2\rangle_{b_\perp} \ ,
\end{equation}
which depend on the centrality of heavy ion collisions. This rapidity dependence can also help to study other magnetic effects in heavy ion collisions, such as the chiral magnetic effects~\cite{Kharzeev:2009pj,Skokov:2016yrj}.

Second, the effects, of course, depend on the magnitude of the magnetic field. More detailed calculations need to formulate the impact dependence of the magnetic fields. In general, as mentioned above, the magnetic field increases from ultra-peripheral to peripheral collisions, but starts to decrease toward more central collisions. 

Third, the direction of transverse momentum kick is correlated to the direction of the magnetic field. Therefore, if this correlation can be measured, we can further distinguish its contribution from other sources for the $p_T$-broadening effects.

It is interesting to note that a recent measurement by the ATLAS collaboration~\cite{ATLAS:2019vxg} shows that there is no strong rapidity dependence on the medium modification of the di-lepton transverse momentum distribution, which will impose a strong constraint on the strength of the magnetic effects discussed above.

\section{Summary and Discussions}

In this paper, we have investigated the electromagnetic productions of di-leptons with very small total transverse momentum in heavy ion collisions, which can provide us a new channel to probe the electromagnetic property of the quark-gluon plasma created in these collisions. In the above theoretical calculations, besides the initial contributions due to the incoming photons, three other important $p_T$-broadening contributions to the acoplanarity of di-leptons measured in the final state are discussed. First, soft photon radiations are taken into account through the Sudakov resummation; Second, the electromagnetic multiple interactions between the lepton pairs and the electric charges inside the quark-gluon plasma medium are resummed incoherently, since these multiple interactions are in general random; Lastly, possible external magnetic fields can be created in heavy ion collisions and these fields may also bring additional transverse momentum broadening.

As an important baseline, we compare the Sudakov effect to the azimuthal angular correlation of the lepton pair in ultra-peripheral collisions measured by the ATLAS collaboration at the LHC. The comparison indicates that this effect is crucial to explain the experimental data in the region with moderately larger acoplanarity. 

It appears that the $p_T$-broadening effects observed by the ATLAS collaboration in the central $PbPb$ collisions suggests the need for additional medium induced transverse momentum broadening coming from the multiple scattering of the leptons in the medium. The acoplanarity of di-lepton pairs can provide us another interesting window to study the properties of quark-gluon plasma. 

The $p_T$-broadening effects due to the possible magnetic fields are considered as well. Through the study on the centrality and rapidity dependence, as well as the correlation with the orientation of the magnetic field, it is possible to distinguish the transverse momentum broadening effects due to the magnetic fields from the multiple scattering effect. We stressed that the broadening coming from the magnetic fields depends on the rapidity difference between the lepton and the anti-lepton. This rapidity dependence can be used to determine the strength of the magnetic field. 

Last but not least, to interpret the $p_T$-broadening phenomena of dilepton productions observed by STAR and ATLAS collaborations as a result of QED interaction of the lepton pair with the medium crucially depend on how precisely we know that the initial state contributions from the incoming photon fluxes of the colliding nuclei. We emphasize that more theoretical developments and experimental measurements are needed to understand this physics. Only with this being resolved, can we reliably apply this process to study the electromagnetic property of the quark-gluon plasma created in heavy ion collisions.

\begin{acknowledgments}
We thank Z.~Xu, W.~Zha, J.~Zhou for discussions and comments. This material is based on work partially supported by the Natural Science Foundation of China (NSFC) under Grant No.~11575070, and by the U.S. Department of Energy, Office of Science, Office of Nuclear Physics, under contract number DE-AC02-05CH11231.
\end{acknowledgments}

\appendix

\section{Azimuthal Angular De-correlation from in the Lab and Lepton Frames}
\label{frame}

In experiments, the azimuthal angular correlation between the final state particles has been applied to study the low total transverse momentum behavior. The two leptons in the final state will be back-to-back in the transverse plane with small angle $\phi_\perp=\pi-\phi_{12}$. We are working at small angle of $\phi_\perp\ll \pi$ region. For convenience, we take one of the leptons' transverse momentum as reference for $-\hat x$ direction, $p_{1\perp}\sim (-|P_T|,0)$, and $p_{2\perp}$ is parameterized by $p_{2T}=(|P_T|\cos\phi_\perp,|P_T|\sin\phi_\perp)$. In the following, we will show that we can compute the azimuthal angular distribution by taking into account additional transverse momentum contribution in the Lab frame or the lepton frame.

\subsection{Lab Frame Calculations}

In the Lab frame, we calculate the azimuthal angular distribution from the total transverse momentum $p_T$ for the lepton pair. This total transverse momentum can come from initial photons' contributions or from the final state medium interaction ($p_T$-broadening).  

We can write down the differential cross section depending on $p_T$,
\begin{equation}
\frac{dN}{d^2 p_T}=\frac{1}{\sigma}\frac{d\sigma}{d^2p_T}=f(p_T) \ ,
\end{equation}
where $f(p_T)$ represents a general form of transverse momentum dependence coming from both incoming photon contribution, soft photon radiation, and the $p_T$-broadening effects in the medium.
Working out the kinematics, we have 
\begin{equation}
\sin(\phi_\perp)=\frac{p_T\sin(\phi')}{P_T}\ ,
\end{equation}
where $\phi'$ is the relative azimuthal angle of $\vec{p}_T$ respect to $\hat x$ direction. The differential cross section can be re-written as
\begin{eqnarray}
&&\frac{dN}{d\phi_\perp}=\int p_T dp_T \left|\frac{P_T\cos(\phi_\perp)}{p_T\cos(\phi')}\right|f(p_T)\nonumber\\
&&~~=\int_{v_q>\sin\phi_\perp} v_qdv_q \frac{\cos\phi_\perp}{\sqrt{v_q^2-\sin^2\phi_\perp}}P_T^2f(p_T)\ ,
\end{eqnarray}
where $v_q=p_T/P_T$. We can further simplify the above expression by making approximations in the correlation limit, i.e., $v_q\ll 1$ and $\phi_\perp\ll 1$, 
\begin{eqnarray}
\frac{dN}{d\phi_\perp}=\frac{1}{2}\int  \frac{dv_q^2}{\sqrt{v_q^2-\phi_\perp^2}}P_T^2f(p_T)\ .
\end{eqnarray}
In the ATLAS experiment, the measurements are presented as functions of the so-called acoplanarity $\alpha$, which is defined as $\alpha=|\phi_\perp|/\pi$. For the $\alpha$ distribution, we have 
\begin{eqnarray}
\frac{dN}{d\alpha}|_{\alpha>0}&=& 2\pi\int \frac{dv_q^2}{\sqrt{v_q^2-\phi_\perp^2}}P_T^2f(p_T)\ .\label{alpha0}
\end{eqnarray}
Let us first assume a simple Gaussian distribution for $f(p_T)$ to check the intuitive $\alpha$ distribution: $f(p_T)=e^{-p_T^2/Q_0^2}/Q_0^2\pi$. With that, we find that 
\begin{eqnarray}
\frac{dN}{d\alpha}|_{\alpha>0}&=& 2\pi\int \frac{dv_q^2}{\sqrt{v_q^2-\phi_\perp^2}}\frac{P_T^2}{\pi Q_0^2}e^{-\frac{P_T^2}{Q_0^2}v_q^2} \nonumber\\ 
&=&\frac{2b_\alpha}{\sqrt{\pi}}e^{-b_\alpha^2\alpha^2}\ ,
\end{eqnarray}
where $b_\alpha=\frac{P_T\pi}{Q_0}$. We can also calculate the average of $\alpha^2$ as
\begin{equation}
\langle \alpha^2\rangle=\frac{1}{2}\frac{Q_0^2}{P_T^2\pi^2} \ ,
\end{equation}
and in terms of $\phi_\perp$, we have 
\begin{equation}
\langle \phi_\perp^2\rangle=\frac{1}{2}\frac{Q_0^2}{P_T^2} \ .
\end{equation}
We have a number of observations. First, it shows that the average angular broadening is, {\it actually}, half of naively expected as the average $p_T$-broadening divided by the lepton's transverse momentum. This will have a significant impact on the interpretation of the experimental measurements on the jet $p_T$-broadening effects from the azimuthal angular correlations in heavy ion collisions. Physically, the jet $p_T$-broadening spreads toward the perpendicular direction respect to the jet direction. However, half of that spreading happens in beam direction in the Lab frame and will not have observable effects on the azimuthal angular correlations. Second, the average of $\langle\alpha^2\rangle$ involves average of $1/P_T^2$, not the average of $P_T$.

In addition, the broadening in azimuthal angular distribution can be model independently related to the additional $p_T$-broadening in the transverse momentum distribution,
\begin{eqnarray}
\Delta \langle \phi_\perp^2\rangle&=&\frac{1}{2}\frac{\Delta \langle \ell_\perp^2\rangle}{P_T^2}=\frac{1}{2}\frac{\Delta \langle p_T^2\rangle}{P_T^2}\ .
\end{eqnarray}
The $p_T$-broadening can be either formulated in the Lab frame as $p_T$ or in the lepton frame as $\ell_\perp$, see, the discussions below. 

\subsection{Lepton Frame Formulation}

It is more convenient to formulate the above results in the lepton frame. It is straightforward to include the medium effects (next section) in this frame, where the leptons are moving along the $\hat z$-direction and additional transverse momentum broadening can be directly formulated. 

If the two leptons are moving in the $\hat z$ direction, we can parameterize the leading lepton (or reference lepton) as $p_1=(E,0,0,E)$ and the away-side lepton as $p_2=(E,\ell_\perp\cos \phi_T, \ell_\perp\sin\phi_T,E\cos\theta_T)$ in the limit of $\ell_\perp\ll E$, where $\ell_\perp=E\sin\theta_T$ represents the imbalance between the two leptons in the final state. Here, we neglect the masses for the leptons since they do not affect the following discussions. The differential cross section can be written as
\begin{equation}
\frac{dN}{d^2\ell_\perp}=\frac{dN}{E^2\sin\theta_T d\sin\theta_T d\phi_T} =f(\ell_\perp)\ ,\label{leptonfe}
\end{equation}
where $\phi_T$ runs from 0 to $2\pi$. However, $\theta_T$ is not directly related to the azimuthal angle between the leptons in the Lab frame. To link to the experimental observables in the Lab frame, we need to work out the details. In the Lab frame, the leptons have non-zero transverse momentum $P_T$ which is same order of $E$, we will parametrize their momenta as $p_1=(E,p_{1T},0,E\cos\theta_0)$ and $p_2=(E,p_{2T}\cos(\pi-\phi_\perp),p_{2T}\sin(\pi-\phi_\perp),p_{2z})$, where $p_{1T}=E\sin\theta_0\sim p_{2T}$ and $\phi_\perp\sim 0$. In the correlation limit, $\vec{p}_{1T}$ and $\vec{p}_{2T}$ are back-to-back with small angular difference $\phi_\perp\ll 1$. From the above kinematics, we will find that
\begin{eqnarray}
&&\sin(\phi_\perp)=\frac{\ell_\perp\sin(\phi_T)}{P_T}\nonumber\\
&&=\frac{E\sin(\theta_T)\sin(\phi_T)}{E\sin\theta_0}=\frac{\sin(\theta_T)\sin(\phi_T)}{\sin\theta_0} \ .
\end{eqnarray}
The differential cross section can be re-written as
\begin{eqnarray}
\frac{dN}{d\phi_\perp}=\int \ell_\perp d\ell_\perp \left|\frac{\sin(\theta_0)\cos(\phi_\perp)}{\sin(\theta_T)\cos(\phi_T)}\right|f(\ell_\perp) \ .
\end{eqnarray}
In the following, we take the approximations in the correlation limit, i.e., $\theta_T\ll 1$ and $\phi_\perp\ll 1$, and find that the $\alpha$ distribution can be written as
\begin{eqnarray}
&&\frac{dN}{d\alpha}|_{\alpha>0}\nonumber\\
&&~~= 4\pi\int \theta_Td\theta_T\frac{\sin(\theta_0)}{\sqrt{\theta_T^2-\sin^2(\theta_0)\phi_\perp^2}}E^2f(\ell_\perp) \ ,
\end{eqnarray}
where $\ell_\perp=E*\theta_T$ and $\alpha$ is define above as $\alpha=|\phi_\perp|/\pi$. With a simple variable change $\theta_T\to \theta_T/\sin(\theta_0)\to v_q$, we can find out that the above equation is identical to that of Eq.~(\ref{alpha0}). All discussions above shall also follow.

\end{document}